\documentclass[twocolumn,tightenlines,prd,aps,showpacs,floatfix]{revtex4}

\def\lsim{\mathrel{\rlap{\lower4pt\hbox{\hskip1pt$\sim$}}
    \raise1pt\hbox{$<$}}}                
\def\gsim{\mathrel{\rlap{\lower4pt\hbox{\hskip1pt$\sim$}}
    \raise1pt\hbox{$>$}}}                
    
\usepackage{graphicx}
\usepackage{color}

\def\sumint{\hbox{$\sum$}\!\!\!\!\!\!\!\int}
\def\square{\vcenter{\vbox{\hrule height.4pt
          \hbox{\vrule width.4pt height8pt
          \kern8pt\vrule width.4pt}\hrule height.4pt}}}

\def\ranglec{\rangle_{\!\!c}}

\newcommand{\beq}{\begin{equation}}
\newcommand{\eeq}{\end{equation}}
\newcommand{\bqa}{\begin{eqnarray}}
\newcommand{\eqa}{\end{eqnarray}}

\begin{document}

\title{Three-loop HTL Free Energy for QED}

\author{Jens O. Andersen}
\affiliation{Department of Physics, Norwegian University of Science
and Technology, H{\o}gskoleringen 5, N-7491 Trondheim, Norway}
\author{Michael Strickland}
\affiliation{Department of Physics, Gettysburg College, Gettysburg, Pennsylvania
17325, USA\\
Frankfurt Institute for Advanced Studies,  D-60438 Frankfurt am Main, Germany}
\author{Nan Su}
\affiliation{Frankfurt Institute for Advanced Studies,  D-60438 Frankfurt 
am Main, Germany}

\date{\today}

\begin{abstract}
We calculate the free energy of a hot gas of electrons and photons
to three loops using the hard-thermal-loop perturbation theory
reorganization of finite-temperature perturbation theory.
We calculate the free energy through three 
loops by expanding in a power series in $m_D/T$, $m_f/T$, 
and $e^2$, where $m_D$ and $m_f$ are
thermal masses and $e$ is the coupling constant. 
We demonstrate that the hard-thermal-loop perturbation reorganization
improves the convergence of the successive approximations
to the QED free energy at large coupling, $e \sim 2$.  The
reorganization is gauge invariant by construction, and 
due to cancellation among various contributions,
we obtain
a completely analytic result for the resummed 
thermodynamic potential at three loops.
Finally, we compare our result with similar calculations that
use the $\Phi$-derivable approach.
\end{abstract}
\maketitle

\section{Introduction}

The thermodynamic functions for hot field theories can be calculated as
a power series in the coupling constant $g$ at weak coupling.
One is primarily interested in calculating the free energy from which
the pressure, energy density, and entropy can be obtained using standard
thermodynamic relations. In the early 1990s the free energy was calculated
to order $g^4$ in Refs.~\cite{wrong,AZ-95} for massless scalar $\phi^4$ theory,
in Ref.~\cite{qed4} for QED and in Ref.~\cite{AZ-95} for non-Abelian gauge 
theories.
The corresponding calculations to order $g^5$ were carried out 
in Refs.~\cite{singh,ea1}, Refs.~\cite{parwani,Andersen} and
Refs.~\cite{KZ-96,BN-96}, respectively.
Recent results have extended the calculation of the QCD free energy
by determining the coefficient of the $g^6 \log(g)$ contribution
\cite{Kajantie:2002wa}.  For massless scalar $\phi^4$ the
perturbative free energy is now known to order $g^6$ \cite{Gynther:2007bw}
and $g^8\log(g)$ \cite{Andersen:2009ct}.

Unfortunately, 
the resulting weak-coupling
approximations, truncated order-by-order in the 
coupling constant, are poorly 
convergent
unless the coupling constant is extremely small. For example, simply comparing
the magnitude of low-order contributions to the $N_f=3$ QCD
free energy one finds that the $g_s^3$ contribution is smaller than the $g_s^2$
contribution only for $g_s \lsim 0.9$ ($\alpha_s \lsim 0.07$).  This is a
troubling situation
since at phenomenologically accessible temperatures near the critical 
temperature 
for the QCD deconfinement phase transition, the strong coupling constant
is on the order of $g_s \sim 2$.  We therefore need methods
which can provide reliable approximations to QCD thermodynamics at
intermediate coupling.

\begin{figure}[t]
\begin{center}
\vspace{1cm}
\includegraphics[width=8.5cm]{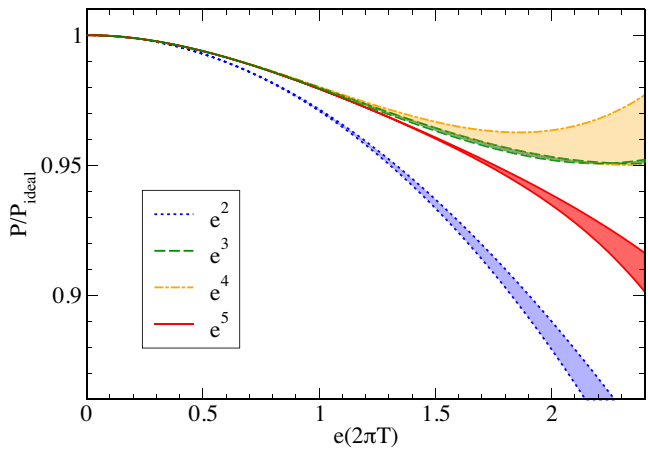}
\end{center}
\vspace{-5mm}
\caption{Successive perturbative approximations to the QED pressure 
(negative of the free energy).  
Each band corresponds to a
truncated weak-coupling expansion accurate to order $e^2$, $e^3$, $e^4$, and 
$e^5$, respectively.  
Shaded bands correspond to variation of the renormalization scale $\mu$ 
between $\pi T$ and $4 \pi T$.}
\label{fig:pertpressure}
\end{figure}

The poor convergence of finite-temperature perturbative expansions of 
the free energy
is not limited to QCD.  The same behavior can be seen in weak-coupling
expansions
in scalar field theory \cite{spt} and QED \cite{parwani,Andersen}.  
In Fig.~\ref{fig:pertpressure} we show the successive perturbative 
approximations
to the QED free energy.  As can be seen from this figure, at couplings larger 
than
$e \sim 1$ the QED weak-coupling approximations also exhibit poor 
convergence.  
For this reason a concerted effort
has been put forth to find a reorganization of finite-temperature 
perturbation theory which converges at phenomenologically relevant
couplings.
Here we will focus on the QED free energy.

There are several ways of systematically 
reorganizing the perturbative expansion to improve its convergence 
and the various approaches have been reviewed 
in Refs.~\cite{birdie,kraemmer, review}. Here we
will focus on the hard-thermal-loop perturbation theory
(HTLpt) method \cite{htl1,AS-01,fermions,htl2,aps1}.  The HTLpt 
method is inspired by variational perturbation 
theory~\cite{yuk,steve,kleinert,deltaexp}.
HTLpt is a gauge-invariant
extension of screened perturbation theory 
(SPT)~\cite{K-P-P-97,CK-98,spt,Andersen:2008bz},
which is a perturbative reorganization for finite-temperature massless scalar 
field theory.
In the SPT approach, one introduces a single variational parameter
which has a simple interpretation as a thermal mass.
In SPT a mass term is added to and subtracted from the scalar Lagrangian,
with the added piece kept as part of the free Lagrangian and the
subtracted piece associated  with the interactions.
The mass parameter is then required to satisfy a variational equation 
which is obtained by the principle of minimal sensitivity.  

This naturally led to the idea that one could apply a similar technique
to gauge theories by adding and subtracting a mass in the Lagrangian.
However, in gauge theories, one cannot simply add and subtract a local mass
term since this would violate gauge invariance.
Instead one adds and subtracts to the Lagrangian a hard-thermal-loop
(HTL) improvement term.
The free part of the Lagrangian then includes the HTL self-energies and 
the remaining terms are treated as perturbations.
Hard-thermal-loop perturbation theory
is a manifestly gauge-invariant approach that can
be applied to static as well as dynamic quantities.
SPT and HTLpt have been applied to four and 
two loops~\cite{spt,Andersen:2008bz,AS-01,htl1,fermions,htl2,aps1}, 
respectively, and convergence is improved
compared to the weak-coupling expansion.

In this paper
we calculate the pressure in QED to three-loop order in HTLpt.  
This will set the stage for the corresponding calculation in full
QCD.  We determine leading-order (LO), next-to-leading-order
(NLO), and next-to-next-to-leading-order (NNLO) expressions
for the HTLpt pressure.  At NNLO the expression is entirely 
analytic and gives a well-defined gap equation (variational equation)
for the electron and photon screening masses.  As we will show,
the NLO and NNLO HTLpt resummed QED free energy give
approximations which show improved convergence for couplings as large
as $e \sim 2.5$ (see Fig.~\ref{fig:NLONNLO}).  In addition, we compare 
our results to those obtained using the 2PI $\Phi$-derivable approach 
\cite{phijm,Borsanyi:2007bf}
and show that at three loops the agreement between the HTLpt and 
$\Phi$-derivable approaches is quite good.

The structure of the paper is as follows.
We give a brief summary of HTLpt in Sec.~II. In Sec.~III, we list
the expressions for the one-, two-, and three-loop diagrams that contribute
to the thermodynamic potential. 
In Sec. IV, we expand the sum-integrals in the mass parameters, and in Sec.~V,
the free energy is calculated. We summarize and draw our conclusions in
Sec.~VI.
In Appendix A, we give the Feynman rules for HTLpt in Minkowski space.
In Appendixes B and C, we
list all sum-integrals and integrals needed in the calculations. 

\section{HTL perturbation theory}

\label{HTLpt}

The Lagrangian density for massless QED in Minkowski space is
\bqa\nonumber
{\cal L}_{\rm QED}&=&
-{1\over4}F_{\mu\nu}F^{\mu\nu}
+i \bar\psi \gamma^\mu D_\mu \psi 
\\&&\hspace{9mm} 
+{\cal L}_{\rm gf}
+{\cal L}_{\rm gh}
+\Delta{\cal L}_{\rm QED}\;.
\label{L-QED}
\eqa
%
Here the field strength is 
$F^{\mu\nu}=\partial^{\mu}A^{\nu}-\partial^{\nu}A^{\mu}$
and the covariant derivative is $D^{\mu}=\partial^{\mu}+ieA^{\mu}$.
The ghost term ${\cal L}_{\rm gh}$ depends on the gauge-fixing term
${\cal L}_{\rm gf}$. In this paper we choose the class of covariant gauges
where the gauge-fixing term is
\bqa
{\cal L}_{\rm gf}&=&-{1\over2\xi}\left(\partial_{\mu}A^{\mu}\right)^2\;,
\eqa
with $\xi$ being the gauge-fixing parameter.
In this class of gauges, the ghost term decouples from the other fields.

The perturbative expansion in powers of $e$
generates ultraviolet divergences.
The renormalizability of perturbative QED guarantees that
all divergences in physical quantities can be removed by
renormalization of the coupling constant $\alpha= e^2/4 \pi$.
There is no need for wavefunction renormalization, because
physical quantities are independent of the normalization of
the field.  There is also no need for renormalization of the gauge
parameter, because physical quantities are independent of the
gauge parameter.

Hard-thermal-loop perturbation theory is a reorganization
of the perturbation
series for thermal gauge theories. In the case of QED, 
the Lagrangian density is written as
\bqa
{\cal L}= \left({\cal L}_{\rm QED}
+ {\cal L}_{\rm HTL} \right) \Big|_{e \to \sqrt{\delta} e}
+ \Delta{\cal L}_{\rm HTL}\;.
\label{L-HTLQCD}
\eqa
The HTL improvement term is
\bqa
{\cal L}_{\rm HTL}=-{1\over2}(1-\delta)m_D^2 
F_{\mu\alpha}\left\langle {y^{\alpha}y^{\beta}\over(y\cdot\partial)^2}
	\right\rangle_{\!\!y}F^{\mu}_{\;\;\beta}
	\nonumber \\
         +(1-\delta)\,i m_f^2 \bar{\psi}\gamma^\mu 
\left\langle {y^{\mu}\over y\cdot D}
	\right\rangle_{\!\!y}\psi
	\, ,
\label{L-HTL}
\eqa
where $y^{\mu}=(1,\hat{{\bf y}})$ is a light-like four-vector,
and $\langle\ldots\rangle_{ y}$
represents an average over the directions
of $\hat{{\bf y}}$.
The term~(\ref{L-HTL}) has the form of the effective Lagrangian
that would be induced by
a rotationally-invariant ensemble of charged sources with infinitely high
momentum. The parameter $m_D$ can be identified with the
Debye screening mass and the parameter $m_f$ can be identified as the
induced finite-temperature electron mass.
HTLpt is defined by treating
$\delta$ as a formal expansion parameter.

The HTL perturbation expansion generates ultraviolet divergences.
In QED perturbation theory, renormalizability constrains the ultraviolet
divergences to have a form that can be cancelled by the counterterm
Lagrangian $\Delta{\cal L}_{\rm QED}$.
We will demonstrate that renormalized perturbation theory can be implemented 
by including a counterterm Lagrangian $\Delta{\cal L}_{\rm HTL}$ among 
the interaction terms in (\ref{L-HTLQCD}).
There is no proof that the HTL perturbation expansion is renormalizable,
so the general structure of the ultraviolet divergences is not known;
however, it was shown in previous papers \cite{htl2,aps1} that it was
possible to renormalize the next-to-leading-order HTLpt prediction for the
free energy of QED using only a vacuum counterterm,
a Debye mass counterterm, and a fermion mass counterterm.  In
this paper we will show that renormalization is also possible at NNLO.

The counterterms necessary are
\bqa
\delta\Delta\alpha&=&N_f{\alpha^2\over3\pi\epsilon}\delta^2\;,
\label{delalpha}
\\ 
\Delta m_D^2&=&N_f\left({\alpha\over3\pi\epsilon}+{\cal O}(\delta^2\alpha^2)
\right)(1-\delta)m_D^2\;,
\label{delmd} \\ 
\Delta m_f^2&=&\left(-{3\alpha\over4\pi\epsilon}+{\cal O}(\delta^2\alpha^2)
\right)
(1-\delta)m_f^2\;,
\label{delmf}\\
\Delta{\cal E}_0&=&\left({1\over128\pi^2\epsilon}+{\cal O}(\delta\alpha)
\right)(1-\delta)^2m_D^4\;.
\label{del1e0}
\eqa

Physical observables are calculated in HTLpt
by expanding them in powers of $\delta$,
truncating at some specified order, and then setting $\delta=1$.
This defines a reorganization of the perturbation series
in which the effects of
the $m_D^2$ and $m_f^2$ terms in~(\ref{L-HTL})
are included to all orders but then systematically subtracted out
at higher orders in perturbation theory
by the $\delta m_D^2$ and $\delta m_f^2$ terms in~(\ref{L-HTL}).
If we set $\delta=1$, the Lagrangian (\ref{L-HTLQCD})
reduces to the QED Lagrangian (\ref{L-QED}).

If the expansion in $\delta$ could be calculated to all orders,
the final result would not depend on $m_D$ or $m_f$ when we set $\delta=1$.
However, any truncation of the expansion in $\delta$ produces results
that depend on $m_D$ and $m_f$.
Some prescription is required to determine $m_D$ and $m_f$
as a function of $T$ and $\alpha$.
We choose to treat both as variational parameters that should be
determined by minimizing the free energy.
If we denote the free energy truncated at some order in $\delta$ by
$\Omega(T,\alpha,m_D,m_f,\mu,\delta)$, our prescription is
\bqa
{\partial \ \ \over \partial m_D}\Omega(T,\alpha,m_D,m_f,\mu,\delta=1) &=& 
0 \, , 
\label{gapmd}\\
{\partial \ \ \over \partial m_f}\Omega(T,\alpha,m_D,m_f,\mu,\delta=1) &=& 
0 \, .
\label{gapmf}
\eqa
%
Since $\Omega(T,\alpha,m_D,m_f,\mu,\delta=1)$ is a function of the
variational parameters $m_D$ and $m_f$, we will refer to it as the
{\it thermodynamic potential}.  We will refer to the variational equations
(\ref{gapmd}) and (\ref{gapmf}) as the {\it gap equations}.  The free energy 
${\cal F}$
is obtained by evaluating the thermodynamic potential at the solution
to the gap equations (\ref{gapmd}) and (\ref{gapmf}).  Other thermodynamic 
functions can then be
obtained by taking appropriate derivatives of ${\cal F}$ with respect to $T$.

\section{Diagrams for the thermodynamic potential}

\begin{figure*}[t]
\begin{center}
\includegraphics[width=7.5cm]{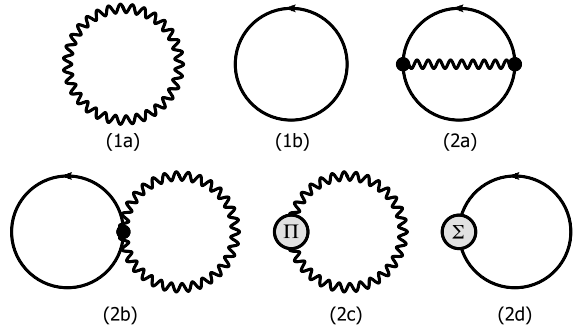}
\end{center}
\caption{Diagrams contributing through NLO in HTLpt.
The undulating lines are photon propagators and the solid lines
are fermion propagators.
A circle with a $\Pi$ indicates a photon self-energy insertion and
a circle with a $\Sigma$ indicates a fermion self-energy insertion.
All propagators and vertices shown are HTL-resummed propagators and vertices.}
\label{fig:dia1}
\end{figure*}

\begin{figure*}[t]
\begin{center}
\includegraphics[width=14.2cm]{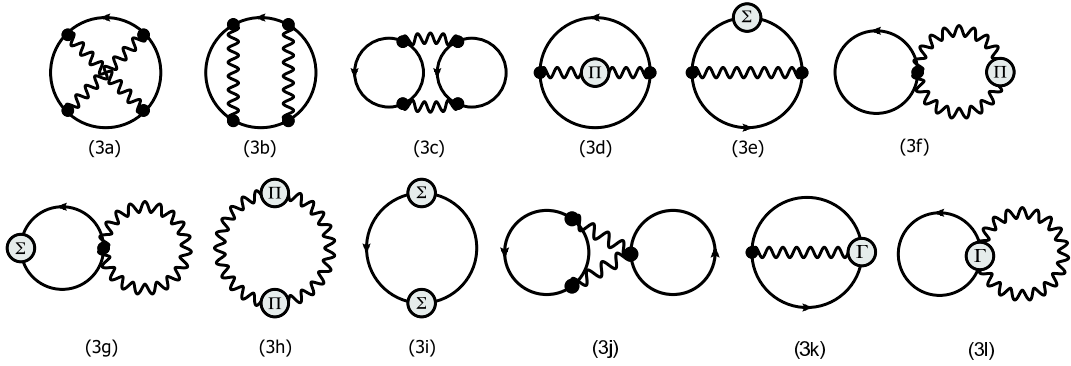}
\end{center}
\caption{Diagrams contributing to NNLO in HTLpt which 
contribute through order $e^5$.
The undulating lines are photon propagators and the solid lines
are fermion propagators.
A circle with a $\Pi$ indicates a photon self-energy insertion and
a circle with a $\Sigma$ indicates a fermion self-energy insertion.
The propagators are HTL-resummed propagators 
and the black dots 
indicate HTL-resummed vertices. The lettered vertices indicate
that only the HTL correction is included.
}
\label{fig:dia2}
\end{figure*}

\begin{figure*}[t]
\begin{center}
\includegraphics[width=11cm]{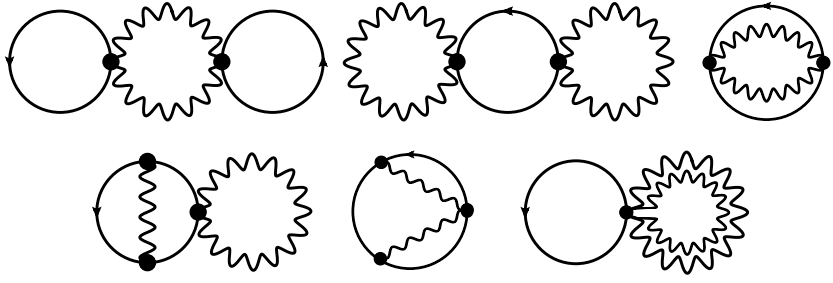}
\end{center}
\caption{Diagrams contributing to NNLO in HTLpt which 
contribute beyond order $e^5$.  The diagrams in the first line above first contribute
at order $e^8$ and the second line at order $e^6$.
The undulating lines are photon propagators and the solid lines
are fermion propagators.
All propagators and vertices shown are HTL-resummed propagators and vertices.}
\label{fig:dia3}
\end{figure*}

In this section, we list the expressions for the diagrams that contribute
to the thermodynamic potential through order $\delta^2$ in HTL
perturbation theory. The diagrams are shown in Figs.~\ref{fig:dia1},
\ref{fig:dia2} and \ref{fig:dia3}.  Because of our dual truncation in $m_D$,
$m_f$, and $e$ the diagrams listed in Fig.~\ref{fig:dia3} do not contribute to
our final expression so we will not explicitly list their integral 
representations.
The expressions here will be given in Euclidean space; however, in Appendix
\ref{app:rules} we present the HTLpt Feynman rules in Minkowski space.

The thermodynamic potential at leading order in HTL perturbation theory for
QED with $N_f$ massless electrons is
\bqa
\Omega_{\rm LO}= {\cal F}_{\rm 1a}
+ N_f{\cal F}_{\rm 1b}+\Delta_0{\cal E}_0\;.
\eqa
Here, ${\cal F}_{1a}$ is the contribution from the photons
\bqa
\!\!\!\!\!\!\!{\cal F}_{1a}\!\!&=&\!\!
-{1\over2}\sumint_{P}\!\left\{
(d-1)\log\left[-\Delta_T(P)\right]+\log\Delta_L(P)
\right\}.
\eqa
The transverse and longitudinal HTL propagators
$\Delta_T(P)$ and $\Delta_L(P)$ 
are given in (\ref{Delta-T}) and (\ref{Delta-L}).
The electron contribution is
\bqa
\label{lq}
{\cal F}_{\rm 1b}=-\sumint_{\{P\}}\log\det\left[P\!\!\!\!/-\Sigma(P)\right]\;.
\eqa
The leading-order vacuum counterterm $\Delta_0{\cal E}_0$ 
is given by
\bqa
\Delta_0{\cal E}_0&=&{1\over128\pi^2\epsilon} m_D^4 \;.
\label{lovac}
\eqa

The thermodynamic potential at 
next-to-leading order (NLO) in HTL perturbation theory
can be written as
\bqa\nonumber
\Omega_{\rm NLO}&=&\Omega_{\rm LO}+
N_f \left({\cal F}_{\rm 2a}+{\cal F}_{\rm 2b}+{\cal F}_{\rm 2d}\right)
 + {\cal F}_{\rm 2c}
+\Delta_1{\cal E}_0
\\&&
+\Delta_1 m_D^2{\partial\over\partial m_D^2}
\Omega_{\rm LO}+\Delta_1 m_f^2{\partial\over\partial m_f^2}\Omega_{\rm LO}\;,
\label{OmegaNLO}
\eqa
%
where $\Delta_1{\cal E}_0$, $\Delta_1m_D^2$, and 
$\Delta_1m_f^2$ are the terms of order
$\delta$ in the vacuum energy density and mass counterterms:
\bqa
\label{dvac1}
\Delta_1{\cal E}_0&=&-{1\over64\pi^2\epsilon}m_D^4\;, \\
\label{dmd1}
\Delta_1m_D^2&=&N_f{\alpha\over3\pi\epsilon}m_D^2\;, \\ 
\Delta_1m_f^2&=&-{3\alpha\over4\pi\epsilon}m_f^2\;.
\label{dmf1}
\eqa
\\
The contributions from the two-loop diagrams
with electron-photon three- and four-point
vertices are
\bqa\nonumber
{\cal F}_{\rm 2a}&=&{1\over2}e^2\sumint_{P\{Q\}}\mbox{Tr}
\left[\Gamma^{\mu}(P,Q,R)S(Q)
\Gamma^{\nu}(P,Q,R)
S(R)\right]
\\ &&
\hspace{2cm}\times\Delta^{\mu\nu}(P)\;,
\label{3qg}
\\  \nonumber
{\cal F}_{\rm 2b}&=&{1\over2}e^2\sumint_{P\{Q\}}\mbox{Tr}
\left[\Gamma^{\mu\nu}(P,-P,Q,Q)S(Q)\right]
\\&&
\hspace{2cm}\times\Delta^{\mu\nu}(P)\;,
\label{4qg}
\eqa
where $R=Q-P$.

The contribution from the HTL photon counterterm diagram 
with a single photon self-energy insertion is
\bqa 
{\cal F}_{\rm 2c}&=&
{1\over2}\sumint_{P}\Pi^{\mu\nu}(P)\Delta^{\mu\nu}(P)\;.
\eqa

The contribution from the HTL electron counterterm diagram with a single
electron self-energy insertion is
\bqa
\label{Sigma}
{\cal F}_{\rm 2d}
&=&-\sumint_{\{P\}}\mbox{Tr}\left[\Sigma(P)S(P)\right]\;.
\eqa
The role of the counterterm diagrams $\rm (2c)$ and $\rm (2d)$ 
is to avoid overcounting
of diagrams when using effective propagators in $\rm (1a)$ and $\rm (1b)$.
Similarly, the role of counterterm diagram $\rm (3k)$
is to avoid overcounting when using effective vertices in $(\rm 2a)$.

The thermodynamic potential at 
next-to-next-to-leading order (NNLO) in HTL perturbation theory
can be written as
\begin{widetext}
\bqa\nonumber
\Omega_{\rm NNLO}&=&\Omega_{\rm NLO}+
N_f^2({\cal F}_{\rm 3c}+{\cal F}_{\rm 3j})
+N_f({\cal F}_{\rm 3a}+{\cal F}_{\rm 3b}
+{\cal F}_{\rm 3d}
+{\cal F}_{\rm 3e}
+{\cal F}_{\rm 3f}+{\cal F}_{\rm 3g}
+{\cal F}_{\rm 3i}+{\cal F}_{\rm 3k}+{\cal F}_{\rm 3l})
+{\cal F}_{\rm 3h}
\\&&\nonumber
+\Delta_2{\cal E}_0
+\Delta_2 m_D^2{\partial\over\partial m_D^2}
\Omega_{\rm LO}+\Delta_2 m_f^2{\partial\over\partial m_f^2}\Omega_{\rm LO}
+\Delta_1 m_D^2{\partial\over\partial m_D^2}
\Omega_{\rm NLO}+\Delta_1 m_f^2{\partial\over\partial m_f^2}\Omega_{\rm NLO}
\\ &&
+{1\over2}\left({\partial^2\over(\partial m_D^2)^2}
\Omega_{\rm LO}\right)\left(\Delta_1m_D^2\right)^2
+{1\over2}\left({\partial^2\over(\partial m_f^2)^2}
\Omega_{\rm LO}\right)\left(\Delta_1m_f^2\right)^2
+{F_{\rm 2a+2b}\over\alpha}\Delta_1\alpha
\;.
\label{OmegaNNLO}
\eqa
where $\Delta_2{\cal E}_0$, $\Delta_2m_D^2$, $\Delta_2m_f^2$, and
$\Delta_1\alpha$
are the terms of order
$\delta^2$ in the vacuum energy density, mass and coupling constant
counterterms:
\bqa
\label{dvac2} 
\Delta_2{\cal E}_0&=&{1\over128\pi^2\epsilon}m_D^4\;, \\
\label{dmd2} 
\Delta_2m_D^2&=&-N_f{\alpha\over3\pi\epsilon}m_D^2\;, \\ 
\Delta_2m_f^2&=&{3\alpha\over4\pi\epsilon}m_f^2\;.
\label{dmf2} 
\eqa

The contributions from the three-loop diagrams are given by
\bqa
\nonumber
{\cal F}_{\rm 3a}
&=&{1\over4}e^4
\sumint_{P\{QR\}}{\rm Tr}\left[\Gamma^{\mu}(-P,Q-P,Q)S(Q)
\Gamma^{\alpha}(Q-R,Q,R)S(R)\Gamma^{\nu}(P,R,R-P)
\right.\\ &&\left.
\times S(R-P)\Gamma^{\beta}(-Q+R,R-P,Q-P)S(Q-P)\right]
\Delta^{\mu\nu}(P)\Delta^{\alpha\beta}(Q-R)\;,
\\ 
\label{ph1}
{\cal F}_{\rm 3b}&=&
{1\over2}e^4\sumint_{P\{QR\}}{\rm Tr}
\left[\Gamma^{\mu}(P,P+Q,Q)S(Q)\Gamma^{\beta}(-R+Q,Q,R)S(R)
\Gamma^{\alpha}(R-Q,R,Q)
\right.\nonumber\\ &&\left.
\times S(Q)\Gamma^{\nu}(-P,Q,P+Q)S(P+Q)\right]
\Delta^{\mu\nu}(P)\Delta^{\alpha\beta}(R-Q)\;,
\\
\label{ph2}
{\cal F}_{\rm 3c}&=&-{1\over4}
e^4\sumint_{P\{QR\}}
{\rm Tr}\left[
\Gamma^{\mu}(P,P+Q,Q)S(Q)\Gamma^{\beta}(-P,Q,P+Q)S(P+Q)
\right]
 \nonumber\\ &&
 \times
{\rm Tr}\left[
\Gamma^{\nu}(-P,R,P+R)S(P+R)
\Gamma^{\alpha}(P,P+R,R)S(R)
\right]\Delta^{\mu\nu}(P)
\Delta^{\alpha\beta}(P)
\;,
\label{ring} \\ \nonumber
{\cal F}_{\rm 3j}&=&-{1\over2}
e^4\sumint_{P\{QR\}}
{\rm Tr}\left[
\Gamma^{\alpha\beta}(P,-P,R,R)
S(R)\right]\Delta^{\alpha\mu}(P)\Delta^{\beta\nu}(P)
\\ 
&&\times
{\rm Tr}\left[
\Gamma^{\mu}(P,P+Q,Q)S(Q)\Gamma^{\nu}(-P,Q,P+Q)S(P+Q)\right]\,.
\eqa
\end{widetext}

The contributions from the two-loop diagrams
with electron-photon three- and four-point
vertices with an insertion of a photon self-energy
\bqa\nonumber
{\cal F}_{\rm 3d}&=&-{1\over2}e^2\sumint_{P\{Q\}}\mbox{Tr}
\left[\Gamma^{\mu}(P,Q,R)S(Q)
\Gamma^{\nu}(P,Q,R)
S(R)\right]
\\ &&\times
\Delta^{\mu\alpha}(P)\Pi^{\alpha\beta}(P)\Delta^{\beta\nu}(P)\;,
\label{3qgpi}
\\   \nonumber
{\cal F}_{\rm 3f}&=&-{1\over2}e^2\sumint_{P\{Q\}}\mbox{Tr}
\left[\Gamma^{\mu\nu}(P,-P,Q,Q)S(Q)\right]
\\&&\hspace{1cm}\times
\Delta^{\mu\alpha}(P)\Pi^{\alpha\beta}(P)\Delta^{\beta\nu}(P)\;,
\label{4qgpi}
\eqa
where $R=Q-P$.

The contributions from the two-loop diagrams
with the electron-photon three and four-point
vertices with an insertion of an electron self-energy are
\bqa\nonumber
{\cal F}_{\rm 3e}&=&-e^2\sumint_{P\{Q\}}
\Delta^{\alpha\beta}(P)
{\rm Tr}\left[
\Gamma^{\alpha}(P,Q,R)S(Q)
\right.\\ &&\left.\times
\Sigma(Q)S(Q)\Gamma^{\beta}(P,Q,R)S(R)
\right]\;,
\label{3qgs}
\\ \nonumber
{\cal F}_{\rm 3g}&=&-{1\over2}e^2
\sumint_{P\{Q\}}
\Delta^{\mu\nu}(P)
\\ &&
\times
{\rm Tr}\left[
\Gamma^{\mu\nu}(P,-P,Q,Q)S(Q)\Sigma(Q)S(Q)
\right]\;,
\label{4qgs}
\eqa
where $R=Q-P$.

The contribution from the HTL photon counterterm diagram 
with two photon self-energy insertions is
\bqa
{\cal F}_{3h}&=&
-{1\over4}\sumint_{P}\Pi^{\mu\nu}(P)\Delta^{\nu\alpha}(P)\Pi^{\alpha\beta}(P)
\Delta^{\beta\mu}(P)\;.
\eqa

The contribution from HTL electron counterterm with two
electron self-energy insertions is
\bqa
{\cal F}_{\rm 3i}&=&{1\over2
}\sumint_{\{P\}}{\rm Tr}\left[S(P)\Sigma(P)S(P)\Sigma(P)
\right]\;.
\label{sigma2}
\eqa
%
The remaining three-loop diagrams involving HTL corrected vertex terms
are given by
\bqa
\nonumber
{\cal F}_{\rm 3k}&=&e^2m_f^2\sumint_{P\{Q\}}\!\!\!\!\!\!\mbox{Tr}
\!\left[\tilde{\cal T}^{\mu}(P,Q,R)S(Q)
\Gamma^{\nu}(P,Q,R)
S(R)\right]
\\ &&\hspace{2cm}\times
\Delta^{\mu\nu}(P)\;,
\label{3k}
\\ \nonumber
{\cal F}_{\rm 3l}&=&-{1\over2}e^2\sumint_{P\{Q\}}\mbox{Tr}
\left[\Gamma^{\mu\nu}(P,-P,Q,Q)S(Q)\right]
\\&&\hspace{2cm}\times
\Delta^{\mu\nu}(P)\;,
\label{3l}
\eqa
where $\tilde{\cal T}^{\mu}$ is the HTL correction term 
given in Eq.~(\ref{T3-def}). Note also that diagram $\rm(3l)$ is the same as
$\rm (2b)$ since there is no tree-level electron-photon four-vertex.

In the remainder of the paper, we work in Landau gauge ($\xi=0$), but
we emphasize that the HTL perturbation theory method of reorganization
is gauge-fixing independent to all orders in $\delta$ (loop expansion) 
by construction.\\

\section{Expansion in the mass parameters}
In the papers~\cite{htl2,aps1}, the free energy was 
reduced to scalar sum-integrals. It was clear that evaluating these
scalar sum-integrals exactly was intractable and the
sum-integrals were calculated approximately 
by expanding them in powers of $m_D/T$ and $m_f/T$.  
We will follow the same strategy in this paper and 
carry out the 
expansion to high enough order  to include all terms 
through  order $e^5$ if $m_D$ and $m_f$ are taken to be of order $e$.
The NLO approximation will be perturbatively accurate to order $e^3$
and the NNLO approximation accurate to order $e^5$.

The free energy can be divided into contributions from hard and soft momenta.
In the one-loop diagrams, the contributions are either hard 
$(h)$ or soft $(s)$, 
while
at the two-loop level, there are hard-hard $(hh)$ and hard-soft $(hs)$ 
contributions.
There are no soft-soft $(ss)$ contributions since one of the loop momenta is
fermionic and always
hard. At three loops there are hard-hard-hard $(hhh)$, 
hard-hard-soft $(hhs)$, and  
hard-soft-soft $(hss)$
contributions. There are no soft-soft-soft $(sss)$ contributions, again due to 
the hard fermionic lines.

In the process of the calculation we will see that there are many cancellations 
between the lower-order HTL-improved diagrams and the higher-order 
HTL-improved counterterm diagrams.  This is by construction and is part of the 
systematic way in which HTLpt converges to the known perturbative expansion.  
For example, one can see that diagrams (2c) and (3h) subtract out the modification 
of the hard gluon propagator due to the HTL-improvement of the propagator in 
diagram (1a).  Likewise, one expects cancellations to occur between diagrams (1b), 
(2d) and (3i); (2a), (3d), (3e) and (3k); and (2b), (3f), (3g), and (3l).  Below we will 
explicitly demonstrate how these cancellations occur.

\subsection{One-loop sum-integrals}

\subsubsection{Hard contribution}
For hard momenta, the self-energies are suppressed by $m_D/T$
and $m_f/T$ relative to the inverse free propagators, so we can expand in powers
of $\Pi_T(P)$, $\Pi_L(P)$, and $\Sigma(P)$.

For the one-loop graph $\rm (1a)$, we need to expand to second order in $m^2_D$:

\begin{widetext}

\bqa\nonumber
{\cal F}_{\rm 1a}^{(h)}&=&{1\over2}(d-1)\sumint_P\log\left(P^2\right)+{1\over2}
m_D^2\sumint_P{1\over P^2}
-{1\over4(d-1)}m_D^4\sumint_P\left[
{1\over P^4}-2{1\over p^2P^2}-2d{1\over p^4}{\cal T}_P
+2{1\over p^2P^2}{\cal T}_P
+d{1\over p^4}\left({\cal T}_P\right)^2
\right]
\\ 
&=&
- {\pi^2 \over 45} T^4
+ {1 \over 24} \left[ 1
        + \left( 2 + 2{\zeta'(-1) \over \zeta(-1)} \right) \epsilon \right]
\left( {\mu \over 4 \pi T} \right)^{2\epsilon} m_D^2 T^2
- {1 \over 128 \pi^2}
\left( {1 \over \epsilon} - 7 + 2 \gamma + {2 \pi^2\over 3} 
\right)
\left( {\mu \over 4 \pi T} \right)^{2\epsilon} m_D^4 \,.
\label{Flo-h}
\eqa
The one-loop graph with a photon self-energy insertion $(\rm 2c$) 
has an explicit factor of $m_D^2$ and so we need only
to expand the sum-integral to first order in $m_D^2$:
\bqa\nonumber
{\cal F}_{\rm 2c}^{(h)}\!\!&=&\!\!-{1\over2}
m_D^2\sumint_P{1\over P^2}
+{1\over2(d-1)}m_D^4\sumint_P\left[
{1\over P^4}-2{1\over p^2P^2}-2d{1\over p^4}{\cal T}_P
+2{1\over p^2P^2}{\cal T}_P
+d{1\over p^4}\left({\cal T}_P\right)^2
\right]\;,\\
\!\!&=&\!\!
-{1 \over 24} \left[ 1
        + \left( 2 + 2{\zeta'(-1) \over \zeta(-1)} \right) \epsilon \right]
\left( {\mu \over 4 \pi T} \right)^{2\epsilon} m_D^2 T^2
+ {1 \over 64 \pi^2}
\left( {1 \over \epsilon} - 7 + 2 \gamma + {2 \pi^2\over 3} 
\right)
\left( {\mu \over 4 \pi T} \right)^{2\epsilon} m_D^4 \,.
\label{ct1}
\eqa

The one-loop graph with two photon self-energy insertions ($\rm 3h$) 
must be expanded to zeroth order in $m_D^2$
\bqa\nonumber
{\cal F}_{\rm 3h}^{(h)}&=&
-{1\over4(d-1)}m_D^4\sumint_P\left[
{1\over P^4}-2{1\over p^2P^2}-2d{1\over p^4}{\cal T}_P
+2{1\over p^2P^2}{\cal T}_P
+d{1\over p^4}\left({\cal T}_P\right)^2
\right]\;.\\
&=&
-{1 \over 128\pi^2}
\left( {1 \over \epsilon} - 7 + 2 \gamma + {2 \pi^2\over 3} 
\right)
\left( {\mu \over 4 \pi T} \right)^{2\epsilon} m_D^4 \,.
\label{ct2}
\eqa
The sum of Eqs.~(\ref{Flo-h})-(\ref{ct2}) is very simple:
\bqa\nonumber
{\cal F}_{\rm 1a+2c+3h}^{(h)}&=&
{1\over2}(d-1)\sumint_P\log\left(P^2\right)\\
&=&-{\pi^2\over45}T^4
\;.
\eqa
This is the free energy of an ideal gas of photons.

The one-loop graph $(1b)$ needs to expanded to second order in $m^2_f$:
\bqa\nonumber
{\cal F}_{\rm 1b}^{(h)}&=&-2\sumint_{\{P\}}\log P^2-4m_f^2
\sumint_{\{P\}}{1\over P^2}
+2m_f^4\sumint_{\{P\}}\left[
{2\over P^4}
-{1\over p^2P^2}+
{2\over p^2P^2}{\cal T}_P
-{1\over p^2P_0^2}\left({\cal T}_P\right)^2
\right]
\\ 
&=&-{7\pi^2\over180}T^4
+{1\over6}
\left[1+\left(2-2\log2+2{\zeta^{\prime}(-1)\over\zeta(-1)}\right)\epsilon
\right]
\left({\mu\over4\pi T}\right)^{2\epsilon}
m_f^2T^2
+{1\over12\pi^2}\left(\pi^2-6\right) 
m_f^4\,.
\label{fc0}
\eqa

The one-loop fermion loop with a fermion self-energy insertion $\rm (2d)$ 
must be expanded to first order in $m_f^2$,
\bqa\nonumber
{\cal F}_{\rm 2d}^{(h)}&=&4m_f^2\sumint_{\{P\}}{1\over P^2}
-4m_f^4\sumint_{\{P\}}\left[
{2\over P^4}-{1\over p^2P^2}
+
{2\over p^2P^2}{\cal T}_P
-{1\over p^2P_0^2}\left({\cal T}_P\right)^2
\right]
\\ &=&
-{1\over6}
\left[1+\left(2-2\log2+2{\zeta^{\prime}(-1)\over\zeta(-1)}\right)\epsilon
\right]
\left({\mu\over4\pi T}\right)^{2\epsilon}
m_f^2T^2
-{1\over6\pi^2}\left(\pi^2-6\right) 
m_f^4\,.
\label{fc1}
\eqa
\end{widetext}
The one-loop fermion loop with two self-energy insertions $\rm(3i)$ must be
expanded to zeroth order in $m_f^2$:
\bqa\nonumber
{\cal F}_{\rm 3i}^{(h)}&=&2m_f^4\sumint_{\{P\}}\left[
{2\over P^4}-{1\over p^2P^2}
+
{2\over p^2P^2}{\cal T}_P
-{1\over p^2P_0^2}\left({\cal T}_P\right)^2
\right]
\\ &=&
{1\over12\pi^2}\left(\pi^2-6\right)
m_f^4\,.
\label{fc2}
\eqa
The sum of Eqs.~(\ref{fc0})-(\ref{fc2}) is particularly simple:
\bqa\nonumber
{\cal F}_{\rm 1b+2d+3i}^{(h)}&=&-2\sumint_{\{P\}}\log P^2 \\
&=&
-{7\pi^2\over180}T^4
\;.
\eqa
This is the free energy of an ideal gas of a single massless fermion.

\subsubsection{Soft contribution}
The soft contributions
in the diagrams $(1a)$, $(2c)$, and $(3h)$
arise from the $P_0=0$ term in the sum-integral.
At soft momentum $P=(0,{\bf p})$, the HTL self-energy functions
reduce to $\Pi_T(P) = 0$ and $\Pi_L(P) = m_D^2$.
The transverse term vanishes in dimensional regularization
because there is no momentum scale in the integral over ${\bf p}$.
Thus the soft contributions come from the longitudinal term only and read
\bqa\nonumber
{\cal F}^{(s)}_{\rm 1a}
&=&{1\over2}T\int_p\log\left(p^2+m_D^2\right)\\
&=&- {m_D^3T\over12\pi}
\left( {\mu \over 2 m} \right)^{2 \epsilon}\left[
1+{8\over3}\epsilon
\right]\;, \nonumber\\
\label{count11}
\\ \nonumber
{\cal F}^{(s)}_{\rm 2c}&=&
-{1\over2}m_D^2T\int_p{1\over p^2+m_D^2}
\\
&=&
{m^3_DT\over 8\pi} \left( {\mu \over 2 m_D} \right)^{2 \epsilon}
\left[1 + 2 \epsilon 
 \right]
\label{count12}
\;, \nonumber \\
\\ \nonumber
{\cal F}^{(s)}_{\rm 3h}&=& - {1\over4}m_D^4T\int_p{1\over(p^2+m_D^2)^2}
\\
&=& - {m^3_DT\over32\pi}
\;.
\eqa
Note that we have kept the terms through order 
$\epsilon$ in Eqs.~(\ref{count11}) and~(\ref{count12})
as they are required in the calculation of the counterterms. 
There is no soft contribution from the leading-order fermion
term~(\ref{lq})
or from the HTL counterterms~(\ref{Sigma}) and (\ref{sigma2}).

\subsection{Two-loop sum-integrals}
For hard momenta, the self-energies are suppressed by $m_D/T$
and $m_f/T$ relative to the inverse free propagators, so we can expand in powers
of $\Pi_T$, $\Pi_L$, and $\Sigma$.

\vspace{6mm}
\subsubsection{(hh) contribution}
The $(hh)$ contribution from~(\ref{3qg}) and~(\ref{4qg}) 
was calculated in Ref.~\cite{aps1} and reads
\begin{widetext}
\bqa\nonumber
{\cal F}_{\rm 2a+2b}^{(hh)}&=&(d-1)e^2\left[\sumint_{\{PQ\}}{1\over P^2Q^2}
-\sumint_{P\{Q\}}{2\over P^2Q^2}\right] 
+2m_D^2e^2\sumint_{P\{Q\}}\left[{1\over p^2P^2Q^2}
{\cal T}_P+{1\over (P^2)^2Q^2}
- {d-2\over d-1}{1\over p^2P^2Q^2}
\right]
\\ \nonumber && 
+m_D^2e^2\sumint_{\{PQ\}}
\left[ {d+1\over d-1}{1\over P^2Q^2r^2}
-{4d\over d-1}{q^2\over P^2Q^2r^4}-{2d\over d-1}
{P\!\cdot\!Q\over P^2Q^2r^4}\right]{\cal T}_R  \\ \nonumber
&& 
+m_D^2e^2\sumint_{\{PQ\}}\left[ {3-d\over d-1}{1\over P^2Q^2R^2}+
{2d\over d-1}{P\!\cdot\! Q\over P^2Q^2r^4}
-{d+2\over d-1}
{1\over P^2Q^2r^2} 
+{4d\over d-1}{q^2\over P^2Q^2r^4}
-{4\over d-1}{q^2\over P^2Q^2r^2R^2} 
\right] \\ \nonumber
&&  
+2m_f^2e^2(d-1)\sumint_{\{PQ\}}\left[ {1\over P^2Q_0^2Q^2}
+{p^2-r^2\over P^2q^2Q_0^2R^2}
\right] {\cal T}_Q
+2m_f^2e^2(d-1)\sumint_{P\{Q\}} \left[{2\over P^2(Q^2)^2}
-{1\over P^2Q_0^2Q^2}{\cal T}_Q\right]
\\ \nonumber
&& 
+2m_f^2e^2(d-1)\sumint_{\{PQ\}}\left[ {d+3\over d-1}{1\over P^2Q^2R^2}
- {2\over P^2(Q^2)^2} 
+{r^2-p^2\over q^2P^2Q^2R^2}\right] \\ \nonumber
& = & {5\pi^2\over72}{\alpha\over\pi}T^4\left[ 1 + \left(3 - {12\over5}\log2 
+4{\zeta'(-1)\over\zeta(-1)}\right)\epsilon\right]
\left({\mu\over4\pi T}\right)^{4\epsilon} \\ \nonumber
&& -{1\over72}\left[{1\over\epsilon} \;+\; 1.30107 
\right]
{\alpha\over\pi}
\left({\mu\over4\pi T}\right)^{4\epsilon}m_D^2T^2
+{1\over8}\left[{1\over\epsilon} \;+\; 8.97544 
\right]
{\alpha\over\pi}
\left({\mu\over4\pi T}\right)^{4\epsilon}m_f^2T^2
\;. \\
&&
\label{33}
\eqa

Consider next the $(hh)$ contribution from~(\ref{3qgpi}) and~(\ref{4qgpi}).
The easiest way to calculate this term, is to expand the two-loop
diagrams $\rm (2a)$ and $(\rm 2b)$ to first order in $m_D^2$. This yields
\bqa\nonumber
{\cal F}_{\rm 3d+3f}^{(hh)}&=&
- 2m_D^2e^2\sumint_{P\{Q\}}
\left[{1\over p^2P^2Q^2}{\cal T}_P+{1\over (P^2)^2Q^2}
- {d-2\over d-1}{1\over p^2P^2Q^2}
\right]
\\ \nonumber && 
- m_D^2e^2\sumint_{\{PQ\}}
\left[ {d+1\over d-1}{1\over P^2Q^2r^2}
-{4d\over d-1}{q^2\over P^2Q^2r^4}-{2d\over d-1}
{P\!\cdot\!Q\over P^2Q^2r^4}\right]{\cal T}_R  
\\ \nonumber
&& 
- m_D^2e^2\sumint_{\{PQ\}}\left[ {3-d\over d-1}{1\over P^2Q^2R^2}+
{2d\over d-1}{P\!\cdot\! Q\over P^2Q^2r^4}
-{d+2\over d-1}
{1\over P^2Q^2r^2} 
+{4d\over d-1}{q^2\over P^2Q^2r^4}
-{4\over d-1}{q^2\over P^2Q^2r^2R^2} 
\right] \\ 
& = & {1\over72}\left[{1\over\epsilon} \;+\; 1.30107 
\right]
{\alpha\over\pi}
\left({\mu\over4\pi T}\right)^{4\epsilon}m_D^2T^2 \;.
\eqa
We also need the $(hh)$ contributions from
the diagrams $(\rm 3e)$, $(\rm 3g)$, $(\rm 3k)$, and $(\rm 3l)$
The first two diagrams are given by~(\ref{3qgs}), ~(\ref{4qgs}), while the
last remaining ones are given by ~(\ref{3k}) and~(\ref{3l}).
The easiest way to calculate these
contributions is to expand the two-loop
diagrams $\rm (2a)$ and $(\rm 2b)$ to first order in $m_f^2$. This yields 
\bqa\nonumber
{\cal F}_{\rm 3e+3g+3k+3l}^{(hh)}&=&
- 2m_f^2e^2(d-1)\sumint_{\{PQ\}}\left[ {1\over P^2Q_0^2Q^2}
+{p^2-r^2\over P^2q^2Q_0^2R^2}
\right] {\cal T}_Q 
-2m_f^2e^2(d-1)\sumint_{P\{Q\}} \left[{2\over P^2(Q^2)^2}
-{1\over P^2Q_0^2Q^2}{\cal T}_Q\right] 
\\ \nonumber 
&& 
- 2m_f^2e^2(d-1)\sumint_{\{PQ\}}\left[ {d+3\over d-1}{1\over P^2Q^2R^2}
- {2\over P^2(Q^2)^2} 
+{r^2-p^2\over q^2P^2Q^2R^2}\right] \\
& = &  - {1\over8}\left[{1\over\epsilon} \;+\; 8.97544 
\right]
{\alpha\over\pi}
\left({\mu\over4\pi T}\right)^{4\epsilon}m_f^2T^2 \;.
\label{44}
\eqa
%
The sum of the terms in~(\ref{33})--(\ref{44}) is very simple
\bqa\nonumber
{\cal F}_{\rm 2a+2b+3d+3e+3f+3g+3k+3l}^{(hh)}&=&
(d-1)e^2\left[\sumint_{\{PQ\}}{1\over P^2Q^2}
-\sumint_{P\{Q\}}{2\over P^2Q^2}\right]
\\ & = & {5 \pi^2\over72} {\alpha\over\pi} T^4\,. 
\eqa
\end{widetext}
This is the two-loop contribution 
from the perturbative expansion of the free energy in QED.

\subsubsection{(hs) contribution}
In the $(hs)$ region, the momentum $P$ is soft. 
The momenta $Q$ and $R$ are always hard. The function that multiplies 
the soft propagator $\Delta_T(0,{\bf p})$, $\Delta_L(0,{\bf p})$,
or $\Delta_X(0,{\bf p})$
can be expanded in powers of the soft momentum ${\bf p}$. 
The soft propagators $\Delta_T(0,{\bf p})$, $\Delta_L(0,{\bf p})$,
and $\Delta_X(0,{\bf p})$ are defined in Eqs.~(\ref{Delta-T:M}), (\ref{Delta-L:M}) and
(\ref{Delta-X}), respectively.
In the case
of $\Delta_T(0,{\bf p})$, the resulting integrals over ${\bf p}$
have no scale and they vanish in dimensional regularization.
The integration measure $\int_{\bf p}$ scales like $m_D^3$,
the soft propagators $\Delta_L(0,{\bf p})$ and
 $\Delta_X(0,{\bf p})$ scale like $1/m_D^2$,
and every power of $p$ in the numerator scales like $m_D$.

The terms that contribute through order $e^2 m_D^3 T$ 
and $e^2m_f^2m_DT$ from ~(\ref{3qg}) and~(\ref{4qg}) 
were calculated in Ref.~\cite{aps1} and read
\begin{widetext}
\bqa\nonumber
{\cal F}_{2a+2b}^{(hs)}&=&2e^2T\int_{p}{1\over p^2+m^2_D}
\sumint_{\{Q\}}\left[
{1\over Q^2}-{2q^2\over Q^4}\right]
+2m_D^2e^2T\int_{p}{1\over p^2+m_D^2}
\sumint_{\{Q\}}
\left[{1\over Q^4}
-{2\over d}(3+d){q^2\over Q^6}+{8\over d}{q^4\over Q^8}
\right]
\\ 
&&
-4m_f^2e^2T\int_{p}{1\over p^2+m_D^2}
\sumint_{\{Q\}}\left[{3\over Q^4}
-{4q^2\over Q^6} -{4\over Q^4} {\cal T}_Q
-{2\over Q^2}\bigg\langle {1\over(Q\!\cdot\!Y)^2} \bigg\rangle_{\!\!\bf \hat y}
\right] \nonumber \\
&=&-{1\over6}\alpha m_DT^3\left[1+\left(3-2\log2
+2{\zeta^{\prime}(-1)\over\zeta(-1)}\right)
\epsilon\right] \nonumber
\left({\mu\over4\pi T}\right)^{2\epsilon}
\left({\mu\over2m_D}\right)^{2\epsilon}
\\&&
+{\alpha\over24\pi^2}\left[{1\over\epsilon}
+\left(1+2\gamma+4\log2\right)
\right]
\left({\mu\over4\pi T}\right)^{2\epsilon}
\left({\mu\over2m_D}\right)^{2\epsilon}
m_D^3T
-{\alpha\over2\pi^2}m_f^2m_DT
\;.
\label{first2}
\eqa

The $(hs)$ contribution from ~(\ref{3qgpi}) and~(\ref{4qgpi}) 
can again be calculated from the 
diagrams $\rm (2a)$ and $(\rm 2b)$ 
by Taylor expanding their contribution 
to first order in $m_D^2$. This yields
\bqa\nonumber
{\cal F}_{\rm 3d+3f}^{(hs)}&=&
2m_D^2e^2T\int_p{1\over(p^2+m_D^2)^2}
\sumint_{\{Q\}}\left[{1\over Q^2}-{2q^2\over Q^4}\right]
-2m_D^2e^2T\int_p{p^2\over(p^2+m^2_D)^2}\sumint_{\{Q\}}
\left[{1\over Q^4}-{2\over d}(3+d){q^2\over Q^6}+{8\over d}{q^4\over Q^8}
\right]
\\ && \nonumber
-4m_D^2m_f^2e^2T\int_p{1\over(p^2+m_D^2)^2}\sumint_{\{Q\}}
\left[{3\over Q^4}
-{4q^2\over Q^6}
-{4\over Q^4}{\cal T}_{Q}-{2\over Q^2}
\bigg\langle
 {1\over(Q\!\cdot\!Y)^2} \bigg\rangle_{\!\!\bf \hat y}
\right] \\ 
& = & {1\over12}\alpha m_DT^3
-{\alpha\over16\pi^2}\left[{1\over\epsilon}
+\left({1\over3}+2\gamma+4\log2\right)
\right]
\left({\mu\over4\pi T}\right)^{2\epsilon}\left({\mu\over2m_D}\right)^{2\epsilon}
m_D^3T
+{\alpha\over4\pi^2}m_f^2m_DT
\;.
\label{hs3d3f}
\eqa

We also need the $(hs)$ contributions from
the diagrams $(\rm 3e)$, $(\rm 3g)$, $(\rm 3k)$, and $(\rm 3l)$
Again we calculate their contributions by
expanding the two-loop
diagrams $\rm (2a)$ and $(\rm 2b)$ to first order in $m_f^2$. 
This yields


\bqa\nonumber
{\cal F}_{3e+3g+3k+3l}^{(hs)}&=&4m_f^2e^2T\int_{p}{1\over p^2+m_D^2}
\sumint_{\{Q\}}\left[{3\over Q^4}
-{4q^2\over Q^6} -{4\over Q^4} {\cal T}_Q
-{2\over Q^2}\bigg\langle {1\over(Q\!\cdot\!Y)^2} \bigg\rangle_{\!\!\bf \hat y}
\right] \\
&=&{\alpha\over2\pi^2}m_f^2m_DT
\;.
\label{last2}
\eqa



\subsubsection{(ss) contribution}
There are no contributions from the $(ss)$ sector since fermionic momenta
are always hard.

\subsection{Three-loop sum-integrals}

\subsubsection{(hhh) contribution}

If all three loop momenta are hard, we can expand the propagators
in powers of $\Pi_{\mu\nu}(P)$ and $\Sigma(P)$.  Through order $e^5$,
we can use bare propagators and vertices.
The diagrams $\rm (3a)$, $\rm (3b)$, and $\rm (3c)$ were calculated in 
Refs.~\cite{qed4,AZ-95}
and their contribution is

\bqa\nonumber
{\cal F}^{(hhh)}_{\rm 3a+3b+3c}&=&{1\over2}(d-1)(d-5)e^4\sumint_{\{PQR\}}
{1\over P^2Q^2R^2(P+Q+R)^2}
\nonumber
 - (d-1)(d-3)e^4
\sumint_{PQ\{R\}}
{1\over P^2Q^2R^2(P+Q+R)^2}
\\
\nonumber
&&+(d-1)^2e^4\sumint_{\{P\}}{1\over P^4}\left[
\sumint_{Q}{1\over Q^2}-\sumint_{\{Q\}}{1\over Q^2}
\right]^2 
+(d-1)^2e^4\sumint_{PQ\{R\}}{1\over P^2Q^2R^2(P+Q+R)^2}
\\ && \nonumber
-2(d-1)^2e^4\sumint_{\{P\}QR}{Q\!\cdot\!R\over P^2Q^2R^2(P+Q)^2(P+R)^2}
-4e^4(d-3)\sumint_{P\{QR\}}{1\over P^4Q^2R^2}
\\ &&
-(d-3)e^4
\sumint_{\{PQR\}}{1\over P^2Q^2R^2(P+Q+R)^2}
-16e^4\sumint_{P\{QR\}}{(Q\!\cdot\!R)^2\over P^4Q^2R^2(P+Q)^2(P+R)^2}\;.
\eqa
Using the expression for the sum-integrals
in the Appendix, we obtain
\bqa\nonumber
{\cal F}_{\rm N_f(3a+3b)+N_f^23c}^{(hhh)}&=&
-N_f^2{5\pi^2\over216}\left({\alpha\over\pi}\right)^2T^4
\left({\mu\over4\pi T}\right)^{6\epsilon}
\left[
{1\over\epsilon}+{31\over10}+{6\over5}\gamma-{192\over25}\log2
\right.
\left.
+{28\over5}{\zeta^{\prime}(-1)\over\zeta(-1)}
-{4\over5}{\zeta^{\prime}(-3)\over\zeta(-3)}\right] 
\\ &&+N_f{\pi^2\over192}\left({\alpha\over\pi}\right)^2T^4\left[35-32\log2
\right]\;.
\label{qedf}
\eqa


\subsubsection{(hhs) contribution}

The diagrams $\rm (3a)$ and $\rm (3b)$ are both 
infrared finite in the limit $m_D\rightarrow0$.
This implies that the $e^5$ contribution is given by 
using a dressed longitudinal propagator and bare vertices.
The ring diagram $\rm (3c)$ is infrared divergent in that limit.
The contribution through $e^5$ is obtained by expanding
in powers of self-energies and vertices. 
Finally, the diagram $\rm (3j)$ also gives a contribution of order $e^5$. Since
the electron-photon four-vertex is already of order $e^2m_f^2$, 
we can use a dressed longitudinal propagator and bare fermion propagators
as well as bare electron-photon three-vertices. 
Note that both $(\rm 3c)$ and $(\rm 3j)$ are proportional to $N_f^2$ and so it
is more convenient to calculate their sum.
One finds
\bqa\nonumber
\nonumber
{\cal F}_{\rm 3a}^{(hhs)}&=&
2(d-1)(d-3)e^4T\int_p{1\over p^2+m_D^2}
\sumint_{\{Q\}}{1\over Q^4}\left[
\sumint_{R}{1\over R^2}-\sumint_{\{R\}}{1\over R^2}
\right] 
\\ && 
+8(d-1)e^4T\int_p{1\over p^2+m_D^2}
\sumint_{Q\{R\}}{q_0r_0\over Q^2R^4(Q+R)^2}\;, \\ &&
\nonumber
\\
{\cal F}_{\rm 3b}^{(hhs)}&=&
-8(d-1)e^4T\int_p{1\over p^2+m_D^2}
\sumint_{Q\{R\}}{q_0r_0\over Q^2R^4(Q+R)^2}\;, \\ \nonumber
{\cal F}_{3c+3j}^{(hhs)}&=&
-4e^4T\int_p{1 \over (p^2+m_D^2)^2}
\left[\sumint_{\{Q\}}
{1\over Q^2}-{2q^2\over Q^4}
\right]^2
\\ && \nonumber
+8e^4T\int_p{p^2 \over (p^2+m_D^2)^2}
\sumint_{\{Q\}}\left[{1\over Q^2}-{2q^2\over Q^4}\right]
\sumint_{\{R\}}\left[
{1\over R^4}-{2\over d}(3+d)
{r^2\over R^6}+{8\over d}{r^4\over R^8}
\right]\\
&&-16
m_f^2e^4T\int_{p}{1\over(p^2+m_D^2)^2}
\sumint_{\{Q\}}\left[{1\over Q^2}-{2q^2\over Q^4}\right]
\sumint_{\{R\}}\left[
{3\over R^4}-{4r^2\over R^6}
-{4\over R^4}{\cal T}_R-{2\over R^2}
\bigg\langle{1\over(R\!\cdot\!Y)^2} \bigg\rangle_{\!\!\bf \hat y}
\right]\;.
\label{hhs3c}
\eqa
%
Using the expressions for the integrals and sum-integrals listed in the
Appendix, we obtain
\bqa
{\cal F}_{\rm N_f(3a+3b)+N_f^2(3c+3j)}^{(hhs)}&=&
-N_f^2{\pi\alpha^2T^5\over18m_D}+N_f^2{\alpha^2m_DT^3\over12\pi}
\left[{1\over\epsilon}+{4\over3}+2\gamma+2\log2+2{\zeta'(-1)\over\zeta(-1)}
\right]\left({\mu\over4\pi T}\right)^{4\epsilon}
\left({\mu\over2m_D}\right)^{2\epsilon} \nonumber \\
&&\hspace{5cm}+N_f{\alpha^2m_DT^3\over4\pi}\ 
-N_f^2{\alpha^2\over3\pi m_D}m_f^2T^3
\;. \nonumber \\
\eqa

\subsubsection{(hss) contribution}

The $(hss)$ modes first start to contribute at order $e^6$, and 
therefore at our truncation order 
the $(hss)$ contributions vanish.  

\subsubsection{(sss) contribution}
There are no contributions from the $(sss)$ sector since fermionic momenta
are always hard.

\section{The Thermodynamic Potential}

In this section we present the final renormalized 
thermodynamic potential explicitly through order
$\delta^2$, aka NNLO.  The final NNLO expression
is completely analytic; however, there are some numerically
determined constants which remain in the final expressions at
NLO.

\subsection{Leading order}

The complete expression for the leading order thermodynamic potential
is given by the sum of
Eqs.~(\ref{Flo-h}),~(\ref{fc0}), and~(\ref{count11}) plus 
the leading vacuum energy counterterm~(\ref{lovac}):
\bqa\nonumber
\Omega_{\rm LO} &=& 
- {\pi^2 T^4\over45}
\left\{ 1 + {7\over4} N_f- {15 \over 2} \hat m_D^2 - 30 
N_f\hat m_f^2
+ 30 \hat m_D^3
+ {45 \over 4}
\left( \log {\hat \mu \over 2}
        - {7\over2} + \gamma + {\pi^2\over 3} \right)
        \hat m_D^4  \nonumber
	- 60 N_f\left(\pi^2-6\right)\hat m_f^4
	\right\} \;.
\\ &&
\label{Omega-LO}
\eqa
where $\hat m_D$, $\hat m_f$, and $\hat \mu$ are dimensionless variables:
\bqa
\hat m_D &=& {m_D \over 2 \pi T}  \;,
\\
\hat m_f &=& {m_f \over 2 \pi T}  \;,
\\
\hat \mu &=& {\mu \over 2 \pi T}  \;. 
\eqa

%

\subsection{Next-to-leading order}
The renormalization contributions at first order in $\delta$ are
\bqa
\Delta_1\Omega&=&\Delta_1{\cal E}_0
+\Delta_1m_D^2{\partial\over\partial m_D^2}\Omega_{\rm LO}+
\Delta_1m_f^2{\partial\over\partial m_f^2}\Omega_{\rm LO}\;.
\eqa
%
Using the results listed in Eqs.~(\ref{dvac1}), (\ref{dmd1}), 
and (\ref{dmf1}), 
the complete contribution from the counterterm at 
first order in $\delta$ is
\bqa
\Delta_1\Omega&=& 
-{\pi^2 T^4\over45} \Bigg\{ {45\over4\epsilon} \hat m_D^4 
	+N_f {\alpha \over \pi} \Bigg[
-{5\over2}
	\left({1\over\epsilon}+2\log{\hat\mu\over 2} + 
2 {\zeta'(-1)\over\zeta(-1)}+2 \right) \hat m_D^2
\nonumber \\ && 
+15
		\left({1\over\epsilon}+2\log{\hat\mu\over 2} - 
2 \log \hat m_D +2 \right) \hat m_D^3
	+ {45 \over 2} 
		\left({1\over\epsilon}+2+2\log{\hat\mu\over 2} - 
2\log2 + 2{\zeta'(-1)\over\zeta(-1)}\right) 
			\hat m_f^2 \Bigg] \Bigg\} \;. 
\label{OmegaVMct1}
\eqa
Adding the NLO counterterms~(\ref{OmegaVMct1}) to the contributions from the
various NLO diagrams, we obtain 
the renormalized NLO thermodynamic potential
\bqa
\Omega_{\rm NLO}&=&
- {\pi^2 T^4\over45} \Bigg\{ 
	1 + {7\over4}N_f - 15 \hat m_D^3 
	- {45\over4}\left(\log\hat{\mu\over2}-{7\over2}+\gamma+{\pi^2\over3}
\right)\hat m_D^4
	+ 60N_f\left(\pi^2-6\right)\hat m_f^4
\nonumber \\ && 
	+ N_f{\alpha\over\pi} \Bigg[ -{25\over8}
	+ 15 \hat m_D
	+5\left(\log{\hat\mu \over 2}-2.33452\right)\hat m_D^2
\nonumber \\ && 
	-45\left(\log{\hat\mu \over 2}+2.19581\right)\hat m_f^2
	- 30\left(\log{\hat\mu \over 2}-{1\over2}
+\gamma+2\log2\right)\!\!\hat m_D^3
	+ 180\hat m_D \hat m_f^2 \Bigg]
\Bigg\} \;.
\label{Omega-NLO}
\eqa
%

\subsection{Next-to-next-to-leading order}

The renormalization contributions at second order in $\delta$ are
\bqa\nonumber
\Delta_2\Omega&=&\Delta_2{\cal E}_0
+\Delta_2m_D^2{\partial\over\partial m_D^2}\Omega_{\rm LO}+
\Delta_2m_f^2{\partial\over\partial m_f^2}\Omega_{\rm LO}
+\Delta_1m_D^2{\partial\over\partial m_D^2}\Omega_{\rm NLO}+
\Delta_1m_f^2{\partial\over\partial m_f^2}\Omega_{\rm NLO}
\\ &&
+{1\over2}\left({\partial^2\over(\partial m_D^2)^2}
\Omega_{\rm LO}\right)\left(\Delta_1m_D^2\right)^2
+{1\over2}\left({\partial^2\over(\partial m_f^2)^2}
\Omega_{\rm LO}\right)\left(\Delta_1m_f^2\right)^2
+{F_{2a+2b}\over\alpha}\Delta_1\alpha
\;.
\eqa
Using the results listed in Eqs.~(\ref{dvac2}), (\ref{dmd2}), 
and (\ref{dmf2}), 
the complete contribution from the counterterms at second order in $\delta$
is
\bqa
\Delta_2\Omega&=& 
-{\pi^2 T^4\over45} \Bigg\{ -{45\over8\epsilon} \hat m_D^4 
	+ N_f {\alpha \over \pi} \Bigg[
{5\over2}
	\left({1\over\epsilon}+2\log{\hat\mu\over 2} + 
2 {\zeta'(-1)\over\zeta(-1)}+2 \right) \hat m_D^2
\nonumber \\ && 
- {45\over2}
		\left({1\over\epsilon}+2\log{\hat\mu\over 2} - 
2 \log \hat m_D +{4\over3} \right) \hat m_D^3
	- {45 \over 2} 
		\left({1\over\epsilon}+2+2\log{\hat\mu\over 2} - 2\log2 + 
2{\zeta'(-1)\over\zeta(-1)}\right) 
			\hat m_f^2 \Bigg] \nonumber \\ && 
	+ N_f^2 \left({\alpha \over \pi}\right)^2 \Bigg[ - {25\over24}
\left({1\over\epsilon} + 4\log{\hat\mu\over 2} + 3 - {12\over5}\log2 + 
4{\zeta'(-1)\over\zeta(-1)}\right) \nonumber \\ && \hspace{20.5mm}
	+{15\over2}
		\left({1\over\epsilon}+4\log{\hat\mu\over 2} - 
2 \log \hat m_D + {7\over3} -2\log2 + 2{\zeta'(-1)\over\zeta(-1)}\right) 
\hat m_D	
	\Bigg]
			\Bigg\} \;.
\label{OmegaVMct2}
\eqa

Adding the NNLO counterterms (\ref{OmegaVMct2}) to the contributions from the 
various NNLO diagrams, we obtain the
renormalized NNLO thermodynamic potential.  We note that at NNLO all
numerically determined subleading coefficients in $\epsilon$ drop out and we 
are left with a final result which
is completely analytic.
The resulting NNLO thermodynamic potential is

\begin{eqnarray}\hspace{-40mm}
\Omega_{\rm NNLO}&=&
- {\pi^2 T^4\over45} \Bigg\{ 
	1 + {7\over4}N_f - {15\over4} \hat m_D^3 
\nonumber \\ &&\hspace{15mm}
	+ N_f {\alpha\over\pi} \Bigg[ -{25\over8}
	+ {15\over2} \hat m_D
	+15 \left(\log{\hat\mu \over 2}-{1\over2}
+\gamma+2\log2\right)\!\!\hat m_D^3
	- 90\hat m_D \hat m_f^2 \Bigg]
\nonumber \\ &&\hspace{15mm}
+ N_f \left({\alpha\over\pi}\right)^2 
\Bigg[{15\over64}(35-32\log2)-{45\over2} \hat m_D\Bigg] 
\nonumber \\ &&\hspace{15mm}
+ N_f^2 \left({\alpha\over\pi}\right)^2 \Bigg[{25\over12}
\left(\log{\hat\mu \over 2}+{1\over20}+{3\over5}\gamma-{66\over25}\log2
+{4\over5}{\zeta^{\prime}(-1)\over\zeta(-1)}
-{2\over5}{\zeta^{\prime}(-3)\over\zeta(-3)}
\right)
\nonumber \\ &&\hspace{40mm}
+{5\over4}{1\over\hat m_D} - 15\left(\log{\hat\mu \over 2}-{1\over2}
+\gamma+2\log2\right)\!\!\hat m_D
+{30}{\hat{m}_f^2\over\hat{m}_D}
\Bigg]
\Bigg\} \;.
\label{Omega-NNLO}
\end{eqnarray}

\end{widetext}

We note that the coupling constant counterterm listed in Eq.~(\ref{delalpha}) 
coincides
with the known one-loop running of the QED coupling constant
\beq
\mu \frac{d e^2}{d \mu} = \frac{N_fe^4}{6 \pi^2} \, .
\label{runningcoupling}
\eeq
Below we will present results as a function of $e$ evaluated at
the renormalization scale $2 \pi T$.  Note that when the free energy is 
evaluated
at a scale different than $\mu = 2 \pi T$ we use Eq.~(\ref{runningcoupling})
to determine the value of the coupling at $\mu = 2 \pi T$.

We have already seen that there are several cancellations that take
place algebraically, irrespective of the values of $m_D$
and $m_f$. For example the $(hh)$ contribution from the two-loop diagrams
($\rm 2a$) and ($\rm 2b$) cancel against the $(hh)$ contribution from the diagrams
($\rm 3d$), ($\rm 3e$), ($\rm 3f$), and ($\rm 3g$). 
As long as only hard momenta are involved,
these cancellations will always take place once the relevant sum-integrals
are expanded in powers of $m_D/T$ and $m_f/T$.
This is no longer the case when soft momenta are involved. However,
further cancellations do take place if one chooses the weak-coupling
values for the mass parameters. 
For example, if one uses the weak-coupling value for the Debye mass,
\bqa\nonumber
m_D^2&=&4 N_f e^2\sumint_{\{Q\}}\left[
{1\over Q^2}-{2q^2\over Q^4}
\right]\\
&=&{4\pi\over3}N_f\alpha T^2\;,
\eqa
the terms proportional to $m_f^2$ in $\Omega_{\rm NNLO}$ cancel 
algebraically
and HTLpt
reduces to the weak-coupling result for the free energy through $e^5$.

\section{Free Energy}

The mass parameters $m_D$ and $m_f$ in hard-thermal-loop 
perturbation theory are in principle completely arbitrary. To complete a 
calculation, it is necessary to specify $m_D$ and $m_f$ as 
functions of $e$ and $T$.  In this section we will
consider two possible mass prescriptions in order to see how much
the results vary given the two different assumptions.  
First we will consider the variational solution resulting from
the thermodynamic potential, Eqs.~(\ref{gapmd}) and (\ref{gapmf}), 
and second we will consider using the
$e^5$ perturbative expansion of the Debye mass 
\cite{Blaizot:1995kg,Andersen} and the $e^3$
perturbative expansion of the fermion mass~\cite{carrington}.


\subsection{Variational Debye mass}

\begin{figure}[t]
\vspace{6mm}
\begin{center}
\includegraphics[width=8.5cm]{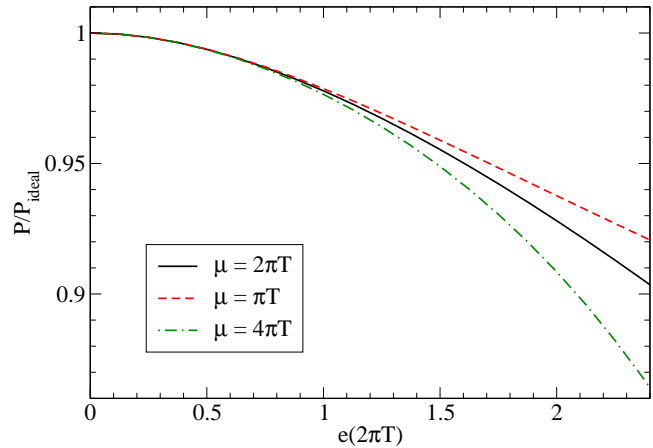}
\end{center}
\caption{NLO HTLpt predictions for the free energy of QED with $N_f=1$ and the 
variational Debye mass. Different curves correspond to varying the 
renormalization scale $\mu$ by a factor of 2 around $\mu=2\pi T$. }
\label{fig:NLO}
\end{figure}

\begin{figure}[t]
\begin{center}
\vspace{9mm}
\includegraphics[width=8.5cm]{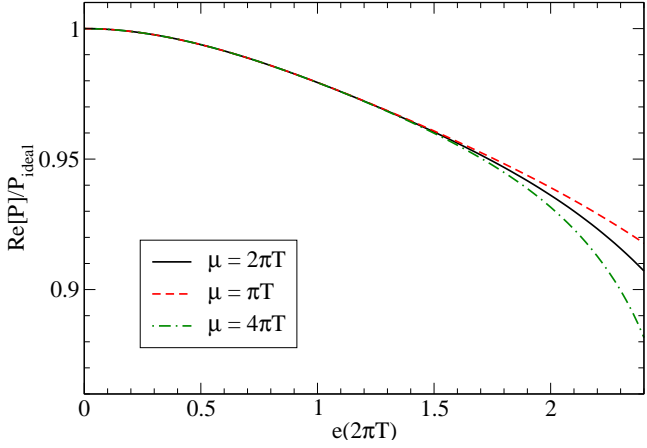}\\
\vspace{9mm}
\includegraphics[width=8.5cm]{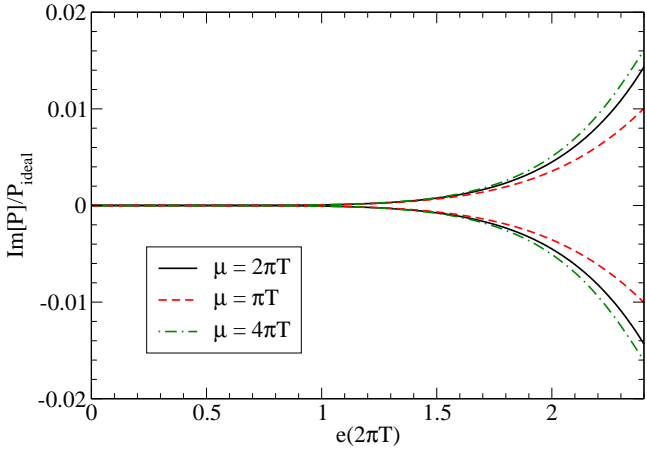}
\end{center}
\caption{Real (top panel) and imaginary (bottom panel) parts of NNLO HTLpt predictions for 
the free energy of QED with $N_f=1$ and the variational Debye mass. Different 
curves correspond to varying the renormalization scale $\mu$ by a factor of 
2 around $\mu=2\pi T$. }

\label{fig:NNLO}
\end{figure}

\begin{figure}[t]
\begin{center}
\vspace{8mm}
\includegraphics[width=8.5cm]{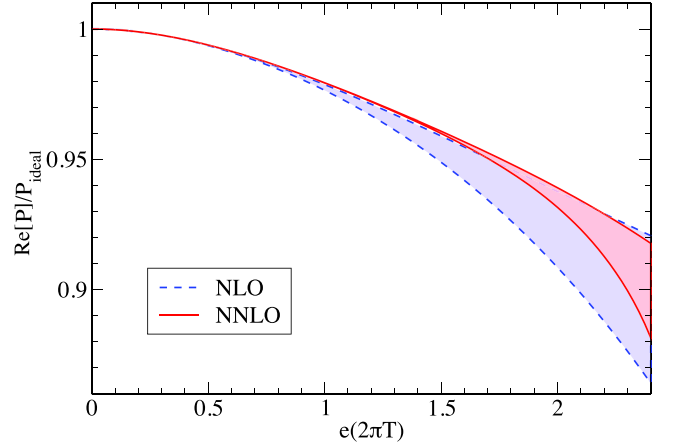}
\end{center}
\caption{A comparison of the renormalization scale variations between NLO and 
NNLO HTLpt predictions for the free energy of QED with $N_f=1$ and the 
variational Debye mass. The bands correspond to varying the renormalization 
scale $\mu$ by a factor of 2 around $\mu=2\pi T$.}
\label{fig:NLONNLO}
\end{figure}

The NLO and NNLO variational Debye mass is determined by solving 
Eqs.~(\ref{gapmd}) and (\ref{gapmf})  
using the NLO and NNLO expressions for the thermodynamic potential, 
respectively. The free energy is then 
obtained by evaluating the NLO and NNLO thermodynamic potentials, 
(\ref{Omega-NLO}) and (\ref{Omega-NNLO}), 
at the solution to the gap equations (\ref{gapmd}) and (\ref{gapmf}). 
Note that at NNLO the gap equation for the
fermion mass is trivial and gives $m_f=0$.
In Figs. \ref{fig:NLO}, \ref{fig:NNLO} and \ref{fig:NLONNLO} we plot 
the NLO and NNLO HTLpt predictions for the free energy of QED with $N_f=1$. 
As can be seen in Fig.~\ref{fig:NLONNLO} 
the renormalization scale variation of the results decreases as one goes from 
NLO to NNLO. 
This is in contrast to weak-coupling expansions for which the scale variation 
can increase as the truncation order is increased.  

One troublesome issue with the variational Debye mass is that at NNLO this 
prescription gives solutions 
for $m_D$ that have a small imaginary part.  We plot the imaginary part of the 
free energy
which results from these imaginary contributions to the variational Debye mass 
in 
Fig. \ref{fig:NNLO} (bottom panel).  The imaginary contributions to the variational 
Debye mass come with both a positive and negative sign
corresponding to the two possible solutions to the quadratic variational gap 
equation.
The positive sign would indicate an unstable solution while the negative sign 
would indicate
a damped solution.  These imaginary parts are most likely an artifact of the 
dual truncation at
order $e^5$; however, without extending the truncation to higher order, it is 
difficult to say.  
They do not occur at NLO in HTLpt in either QED or QCD.  We note that a 
similar effect 
has also been observed in NNLO
screened perturbation theory in scalar theories \cite{spt}.  Because of this 
complication, in the next 
subsection we will discuss a different mass prescription in order to assess 
the impact of these
small imaginary parts.

\subsection{Perturbative Debye and fermion masses}

\begin{figure}[t]
\begin{center}
\vspace{5mm}
\includegraphics[width=8.5cm]{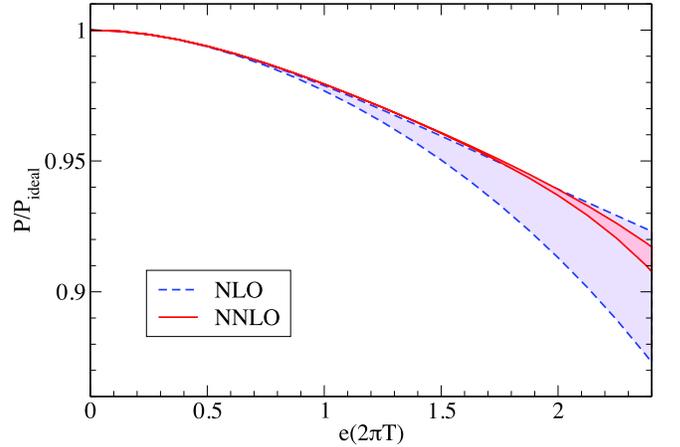}
\end{center}
\caption{A comparison of the renormalization scale variations between NLO and 
NNLO HTLpt predictions for the free energy of QED with $N_f=1$ using the 
perturbative thermal masses given in Eqs.~(\ref{mass})  and (\ref{fmass}). 
The bands correspond to varying the renormalization 
scale $\mu$ by a factor of 2 around $\mu=2\pi T$.}
\label{fig:NLONNLOpmd}
\end{figure}

The perturbative Debye and fermion masses 
for QED have been calculated  through order 
$e^5$~\cite{Blaizot:1995kg,Andersen} and $e^3$~\cite{carrington}, respectively:
\begin{eqnarray}
\label{mass}
m_D^{2}&=&{1\over3}N_fe^2T^{2}\Big [1
-\frac{e^{2}}{24\pi^{2}}\left(4\gamma + 7  
+4\log \frac{\hat\mu}{2}+8\log2\right) \nonumber \\ 
&&\hspace{1.5cm} +{e^3\sqrt{3}\over4\pi^3}\Big] \; ,\\
m_f^2&=&{1\over8}N_fe^2T^2\left[
1-{2.854\over4\pi}e\right]\;.
\label{fmass}
\end{eqnarray}
Plugging~(\ref{mass})  and~(\ref{fmass}) 
into the NLO and NNLO thermodynamic potentials, 
(\ref{Omega-NLO}) and (\ref{Omega-NNLO}), we obtain the results shown in 
Fig.~\ref{fig:NLONNLOpmd}. The renormalization scale variation is quite 
small in the NNLO result. 

\subsection{Comparison with the $\Phi$-derivable approach}

\begin{figure}[t]
\begin{center}
\vspace{6mm}
\includegraphics[width=8.5cm]{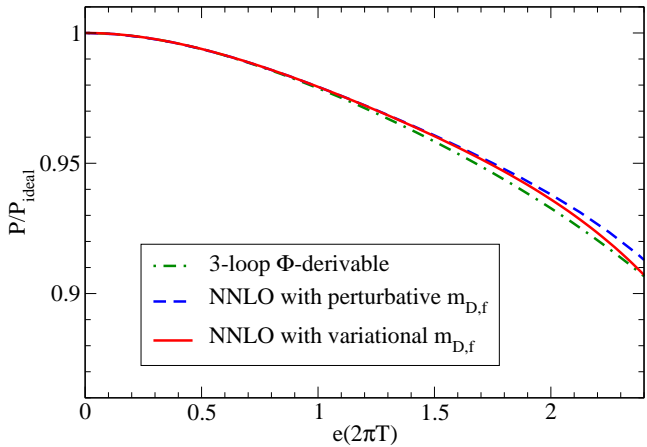}
\end{center}
\caption{A comparison of the predictions for the free energy of QED with 
$N_f=1$ between three-loop $\Phi$-derivable approximation \cite{phijm} and NNLO 
HTLpt at $\mu=2\pi T$.}
\label{fig:PhivsNNLO}
\end{figure}

Having obtained the NNLO HTLpt result for the free energy we can now compare 
the
results obtained using this reorganization with results obtained 
within the $\Phi$-derivable
approach.  In Fig. \ref{fig:PhivsNNLO} we show a comparison of our NNLO HTLpt 
results 
with a three-loop calculation obtained previously using a truncated three-loop 
$\Phi$-derivable approximation
\cite{phijm}.  For the NNLO HTLpt prediction we show the results obtained 
using both the
variational and perturbative mass prescriptions.  As can be seen from this 
figure, there is
very good agreement between the NNLO $\Phi$-derivable and HTLpt approaches 
out to large
coupling.  In all cases we have chosen the renormalization scale to be 
$\mu = 2\pi T$.

\begin{figure}[t]
\begin{center}
\vspace{6mm}
\includegraphics[width=8.5cm]{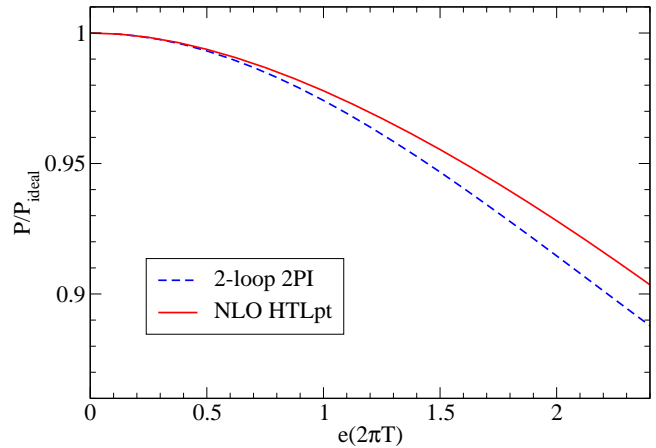}
\end{center}
\caption{A comparison of the predictions for the free energy of QED with 
$N_f=1$ between the two-loop 2PI approximation in Landau gauge \cite{Borsanyi:2007bf}
and NLO HTLpt at $\mu=2\pi T$.}
\label{fig:2PIvsNLO}
\end{figure}

As a further consistency check, in Fig.~\ref{fig:2PIvsNLO} we show a 
comparison between
the untruncated two-loop numerical $\Phi$-derivable approach calculation of 
Ref.~\cite{Borsanyi:2007bf}
and our NLO HTLpt result using the variational mass.  In both cases we have 
chosen the 
renormalization scale to be $\mu = 2\pi T$.  From this figure we see that 
there is a reasonable
agreement between the NLO numerical $\Phi$-derivable and NLO HTLpt results; 
however,
the agreement is not as good as the corresponding NNLO results.  

We note that the results of \cite{Borsanyi:2007bf} were computed
in the Landau gauge ($\xi=0$).  As detailed in their paper, their result is 
gauge dependent.
Such gauge dependence is unavoidable
in the 2PI $\Phi$-derivable approach since it only uses dressed propagators.  
In Ref.~\cite{phijm}
it was explicitly shown that the two-loop $\Phi$-derivable Debye mass is 
gauge independent 
only up to order $e^2$, resulting in gauge variation of the free energy at 
order $e^4$.  
This is in agreement with general theorems stating that the gauge variance 
appears at one
order higher than the truncation \cite{Arrizabalaga:2002hn}.

\section{Conclusions}

In this paper we calculated the three-loop HTLpt thermodynamic potential in 
QED.  
Having obtained this we applied two mass prescriptions, variational and 
perturbative, to 
fix the {\em a priori} undetermined parameters $m_D$ and $m_f$ that appear in 
the HTL-improved
Lagrangian.  We found that the resulting expressions for the free energy 
were the same
to an accuracy of 0.6\% at $e=2.4$ giving us confidence in the prediction. 
We also
compared the HTLpt three-loop result with a three-loop 
$\Phi$-derivable approach \cite{phijm} and found agreement at the 
subpercentage level at large coupling.  

In addition, we showed that the HTLpt NLO and NNLO approximations have improved
convergence at large coupling compared to the naively truncated weak-coupling 
expansion
and that the renormalization scale variation at NNLO using both the 
variational and perturbative mass
prescriptions was quite small. Therefore, the NNLO HTLpt method result seems 
to be quite reliable.
This is important since, unlike the $\Phi$-derivable approach, the HTLpt 
reorganization 
is gauge invariant by construction and is formulated directly in Minkowski 
space allowing it to, 
in principle, also be applied to the calculation of dynamical quantities.  

The renormalization of the three-loop thermodynamic potential required only 
known 
vacuum, mass, and coupling constant counterterms, and the resulting running 
coupling was 
found to coincide with the QED one-loop running.  This provides further evidence 
that the
HTLpt framework is renormalizable despite the new divergences which are 
introduced
during HTL improvement.

Finally, we note that at three loops we could obtain an entirely analytic 
expression for the 
renormalized NNLO thermodynamic potential.  There were a number of cancellations
that took place during renormalization which resulted in an expression that 
was
independent of any numerically determined subleading coefficients in the 
sum-integrals.
We expect similar cancellations to also occur in non-Abelian gauge theories 
which will
greatly simplify the calculation.  Computing the three-loop HTLpt reorganized 
free energy
for QCD is in progress.

\section*{Acknowledgments}

We thank H. St\"ocker for his encouragement and support for this endeavor.
We thank S. Borsanyi and U. Reinosa for providing us with predictions for the QED 
pressure
from their calculation.
N. S. thanks  J.W. Qiu, J.Y. Jia, R. Lacey, the Physics Department at 
Gettysburg College,
and the Niels
Bohr Institute for hospitality.
N. S. was supported by the Frankfurt International Graduate School for Science. 
M. S. was supported in part 
by the Helmholtz International Center for FAIR Landesoffensive zur Entwicklung 
Wissenschaftlich-\"Okonomischer Exzellenz program.
J. O. A. thanks the Niels Bohr International Academy and the Niels
Bohr Institute for kind hospitality during the final stages of this project.

\appendix

\section{HTL Feynman Rules}
\setcounter{equation}{0}
\label{app:rules}
In this appendix, we present Feynman rules for HTL perturbation theory 
in QED.  We give explicit expressions for the propagators 
and for the electron-photon three- and four-vertices.
The Feynman rules are given in Minkowski space to facilitate 
applications to real-time processes.
A Minkowski momentum is denoted $p = (p_0, {\bf p})$,
and the inner product is $p \cdot q = p_0 q_0 - {\bf p} \cdot {\bf q}$.
The vector that specifies the thermal rest frame 
is $n = (1,{\bf 0})$.

\subsection{Photon self-energy}

The HTL photon self-energy tensor for a photon of momentum $p$ is
\bqa
\label{a1}
\Pi^{\mu\nu}(p)=m_D^2\left[
{\cal T}^{\mu\nu}(p,-p)-n^{\mu}n^{\nu}
\right]\;.
\eqa
%
The tensor ${\cal T}^{\mu\nu}(p,q)$, which is defined only for momenta
that satisfy $p+q=0$, is
\bqa
{\cal T}^{\mu\nu}(p,-p)=
\left \langle y^{\mu}y^{\nu}{p\!\cdot\!n\over p\!\cdot\!y}
\right\rangle_{\bf\hat{y}} \;.
\label{T2-def}
\eqa
%
The angular brackets indicate averaging
over the spatial directions of the lightlike vector $y=(1,\hat{\bf y})$.
The tensor ${\cal T}^{\mu\nu}$ is symmetric in $\mu$ and $\nu$
and satisfies the ``Ward identity''
\bqa
p_{\mu}{\cal T}^{\mu\nu}(p,-p)=
p\!\cdot\!n\;n^{\nu}\;.
\label{ward-t2}
\eqa
%
The self-energy tensor $\Pi^{\mu\nu}$ is therefore also
symmetric in $\mu$ and $\nu$ and satisfies
\bqa
p_{\mu}\Pi^{\mu\nu}(p)&=&0\;,\\
\label{contr}
g_{\mu\nu}\Pi^{\mu\nu}(p)&=&-m_D^2\;.
\eqa
%

The photon self-energy tensor can be expressed in terms of two scalar functions,
the transverse and longitudinal self-energies $\Pi_T$ and $\Pi_L$,
defined by
\bqa
\label{pit2}
\Pi_T(p)&=&{1\over d-1}\left(
\delta^{ij}-\hat{p}^i\hat{p}^j
\right)\Pi^{ij}(p)\;, \\
\label{pil2}
\Pi_L(p)&=&-\Pi^{00}(p)\;,
\eqa
%
where ${\bf \hat p}$ is the unit vector
in the direction of ${\bf p}$.
In terms of these functions, the self-energy tensor is
\bqa
\label{pi-def}
\Pi^{\mu\nu}(p) \;=\; - \Pi_T(p) T_p^{\mu\nu}
- {1\over n_p^2} \Pi_L(p) L_p^{\mu\nu}\;,
\eqa
%
where the tensors $T_p$ and $L_p$ are
\bqa
T_p^{\mu\nu}&=&g^{\mu\nu} - {p^{\mu}p^{\nu} \over p^2}
-{n_p^{\mu}n_p^{\nu}\over n_p^2}\;,\\
L_p^{\mu\nu}&=&{n_p^{\mu}n_p^{\nu} \over n_p^2}\;.
\eqa
%
The four-vector $n_p^{\mu}$ is
\bqa
n_p^{\mu} \;=\; n^{\mu} - {n\!\cdot\!p\over p^2} p^{\mu}
\eqa
%
and satisfies $p\!\cdot\!n_p=0$ and $n^2_p = 1 - (n\!\cdot\!p)^2/p^2$.
Equation~(\ref{contr}) reduces to the identity
\bqa
(d-1)\Pi_T(p)+{1\over n^2_p}\Pi_L(p) \;=\; m_D^2 \;.
\label{PiTL-id}
\eqa
%
We can express both self-energy functions in terms of the function
${\cal T}^{00}$ defined by (\ref{T2-def}):
\bqa
\Pi_T(p)&=& {m_D^2 \over (d-1) n_p^2}
\left[ {\cal T}^{00}(p,-p) - 1 + n_p^2  \right] ,
\label{PiT-T}
\\
\Pi_L(p)&=& m_D^2
\left[ 1- {\cal T}^{00}(p,-p) \right] ,
\label{PiT-L}
\eqa
%

In the tensor ${\cal T}^{\mu \nu}(p,-p)$ defined in~(\ref{T2-def}),
the angular brackets indicate the angular average over
the unit vector $\hat{\bf y}$.
In almost all previous work, the angular average in~(\ref{T2-def}) has been
taken in $d=3$ dimensions. For consistency of higher-order radiative
corrections, it is essential to take the angular average in $d=3-2\epsilon$
dimensions and analytically continue to $d=3$ only after all poles in
$\epsilon$ have been cancelled.
Expressing the angular average as an integral over the cosine of an angle,
the expression for the $00$ component of the tensor is
\bqa
\!\!\!{\cal T}^{00}(p,-p) \!\! &=& \!\! {w(\epsilon)\over2} \!\!
\int_{-1}^1 \!\! dc\;(1-c^2)^{-\epsilon}{p_0\over p_0-|{\bf p}|c} \, ,
\label{T00-int}
\eqa
%
where the weight function $w(\epsilon)$ is
\bqa
w(\epsilon)={\Gamma(2-2\epsilon)\over\Gamma^2(1-\epsilon)}2^{2\epsilon}
= {\Gamma({3\over2}-\epsilon)
        \over \Gamma({3\over2}) \Gamma(1-\epsilon)} \;.
\label{weight}
\eqa
%
The integral in (\ref{T00-int}) must be defined so that it is analytic
at $p_0=\infty$.
It then has a branch cut running from $p_0=-|{\bf p}|$ to $p_0=+|{\bf p}|$.
If we take the limit $\epsilon\rightarrow 0$, it reduces to
\begin{eqnarray}
{\cal T}^{00}(p,-p) &=&
{p_0 \over 2|{\bf p}|}
                \log {p_0 +|{\bf p}| \over p_0-|{\bf p}|}\;,
\end{eqnarray}
%
which is the expression that
appears in the usual HTL self-energy functions.

\label{app:prop}

The Feynman rule for the photon propagator is
\bqa
i \Delta_{\mu\nu}(p) \;,
\eqa
%
where the photon propagator tensor $\Delta_{\mu\nu}$
depends on the choice of gauge fixing.
We consider two possibilities that introduce an arbitrary
gauge parameter $\xi$:  general covariant gauge and
general Coulomb gauge.
In both cases, the inverse propagator reduces in the
limit $\xi\rightarrow\infty$ to
\bqa
\Delta^{-1}_{\infty}(p)^{\mu\nu}&=&
-p^2 g^{\mu \nu} + p^\mu p^\nu - \Pi^{\mu\nu}(p)\;.
\label{delta-inv:inf0}
\eqa
%
This can also be written
\bqa
\Delta^{-1}_{\infty}(p)^{\mu\nu}&=&
- {1 \over \Delta_T(p)}       T_p^{\mu\nu}
+ {1 \over n_p^2 \Delta_L(p)} L_p^{\mu\nu}\;,
\label{delta-inv:inf}
\eqa
%
where $\Delta_T$ and $\Delta_L$ are the transverse and longitudinal
propagators:
\bqa
\Delta_T(p)&=&{1 \over p^2-\Pi_T(p)}\;,
\label{Delta-T:M}
\\
\Delta_L(p)&=&{1 \over - n_p^2 p^2+\Pi_L(p)}\;.
\label{Delta-L:M}
\eqa
%
The inverse propagator for general $\xi$ is
\bqa
\Delta^{-1}(p)^{\mu\nu}&=&
\Delta^{-1}_{\infty}(p)^{\mu\nu}-{1\over\xi}
p^{\mu}p^{\nu}\hspace{0.2cm}\mbox{covariant}\;,
\label{Delinv:cov}
\\ \nonumber
&=&\Delta^{-1}_{\infty}(p)^{\mu\nu}-{1\over\xi}
\left(p^{\mu}-p\!\cdot\!n\;n^{\mu}\right)
\left(p^{\nu}-p\!\cdot\!n\;n^{\nu}\right)
\\ && 
\hspace{3.7cm}\mbox{Coulomb}\;.
\label{Delinv:C}
\eqa
%
The propagators obtained by inverting the tensors in~(\ref{Delinv:C})
and~(\ref{Delinv:cov}) are
\bqa
\Delta^{\mu\nu}(p)&=&-\Delta_T(p)T_p^{\mu\nu}
+\Delta_L(p)n_p^{\mu}n_p^{\nu}
- \xi {p^{\mu}p^{\nu} \over (p^2)^2}
\nonumber \\
&& \hspace{3.7cm}\mbox{covariant}\;,
\label{D-cov}
\\
&=&-\Delta_T(p)T_p^{\mu\nu}
+\Delta_L(p)n^{\mu}n^{\nu}-\xi{p^{\mu}p^{\nu}\over\left(n_p^2p^2\right)^2}
\nonumber \\
&& \hspace{3.7cm}
\mbox{Coulomb}\;.
\label{D-C}
\eqa
%

It is convenient to define the following combination of propagators:
\bqa
\Delta_X(p) &=& \Delta_L(p)+{1\over n_p^2}\Delta_T(p) \;.
\label{Delta-X}
\eqa
%
Using (\ref{PiTL-id}), (\ref{Delta-T:M}), and (\ref{Delta-L:M}),
it can be expressed in the alternative form
\bqa
\Delta_X(p) &=&
\left[ m_D^2 - d \, \Pi_T(p) \right] \Delta_L(p) \Delta_T(p) \;,
\label{Delta-X:2}
\eqa
%
which shows that it vanishes in the limit $m_D \to 0$.
In the covariant gauge, the propagator tensor can be written
\bqa\nonumber
\Delta^{\mu\nu}(p) &=&
\left[ - \Delta_T(p) g^{\mu \nu} + \Delta_X(p) n^\mu n^\nu \right]
\\ \nonumber&&
- {n \!\cdot\! p \over p^2} \Delta_X(p)
        \left( p^\mu n^\nu  + n^\mu p^\nu \right)
\\
&&
+ \left[ \Delta_T(p) + {(n \!\cdot\! p)^2 \over p^2} \Delta_X(p)
        - {\xi \over p^2} \right] {p^\mu p^\nu \over p^2} \;. \nonumber \\
\label{gprop-TC}
\eqa
%
This decomposition of the propagator into three terms
has proved to be particularly convenient for explicit calculations.
For example, the first term satisfies the identity
\bqa
\nonumber
\left[- \Delta_T(p) g_{\mu \nu} + \Delta_X(p) n_\mu n_\nu \right]
\Delta^{-1}_{\infty}(p)^{\nu\lambda}  &=&
\\
&& \hspace{-5cm}
{g_\mu}^\lambda - {p_\mu p^\lambda \over p^2}
+ {n \!\cdot\! p \over n_p^2 p^2} {\Delta_X(p) \over \Delta_L(p)}
        p_\mu n_p^\lambda \;.
\label{propid:2}
\eqa
%
\subsection{Electron self-energy}
The HTL self-energy of an electron with momentum $p$ is given by
\bqa
\label{selfq}
\Sigma(P)=m_f^2/\!\!\!\!{\cal T}(p)
\;,
\eqa
where
\bqa
\label{deftf}
{\cal T}^{\mu}(p)=
\left\langle{y^{\mu}\over p\cdot y}
\right\rangle_{\hat{\bf y}}
\;.
\eqa
Expressing the angular average as an integral over the cosine of an angle,
the expression is
\bqa 
\label{def-tf}
{\cal T}^{\mu}(p)={w(\epsilon)\over2}
\int_{-1}^1dc\;(1-c^2)^{-\epsilon}{y^{\mu}\over p_0-|{\bf p}|c}\;,
\eqa
The integral in (\ref{def-tf}) must be defined so that it is analytic 
at $p_0=\infty$.
It then has a branch cut running from $p_0=-|{\bf p}|$ to $p_0=+|{\bf p}|$.  
In three dimensions, this reduces to
\bqa\nonumber
\Sigma(P)&=&
{m_f^2\over 2|{\bf p}|}\gamma_0\log{p_0+|{\bf p}|\over p_0-|{\bf p|}}
\\&&\!\!\!
+{m_f^2\over |{\bf p}|}{\bf \gamma}\cdot \hat{\bf p}
\left(1-{p_0\over 2|{\bf p}|}\log{p_0+|{\bf p}|\over p_0-|{\bf p|}}\right) .
\eqa

\subsection{Electron propagator}
The Feynman rule for the electron propagator is 
\bqa
iS(p)\;.
\eqa
The electron propagator can be written as
\bqa
\label{qprop}
S(p)={1\over/\!\!\!p-\Sigma(p)}\;,
\eqa
where the electron self-energy is given by~(\ref{selfq}).
The inverse electron propagator can be written as
\bqa
S^{-1}(p)=/\!\!\!p-\Sigma(p)\;.
\eqa
This can be written as
\bqa
S^{-1}(p)=/\!\!\!\!{\cal A}(p)\;,
\eqa
where we have organized $A_0(p)$ and $A_S(p)$ into:
\bqa
\label{qself}
A_{\mu}(p)=(A_0(p),A_S(p)\hat{\bf p})\;.
\eqa
The functions $A_0(p)$ and $A_S(p)$ are defined
\bqa
\label{aodef}
A_0(p)&=&p_0-{m_f^2\over p_0}{\cal T}_p\;,\\
A_S(p)&=& |{\bf p}|+{m_f^2\over |{\bf p}|}\left[1-{\cal T}_p\right]\;.
\label{asdef}
\eqa

\subsection{Electron-photon vertex}
The electron-photon vertex with outgoing photon momentum $p$, 
incoming electron momentum $q$, outgoing electron momentum $r$, 
and Lorentz index $\mu$ is
\bqa
\label{3qgv}
\Gamma^{\mu}(p,q,r)
=e\left(\gamma^{\mu}-m_f^2\tilde{{\cal T}}^{\mu}(p,q,r)\right)\;.
\eqa
The tensor in the HTL correction term is only defined for $p-q+r=0$:
\bqa
\tilde{{\cal T}}^{\mu}(p,q,r)
=\left\langle
y^{\mu}\left({y\!\!\!/\over q\!\cdot\!y\;\;r\!\cdot\!y}\right)
\right\rangle_{\hat{\bf y}}.
\label{T3-def}
\eqa
This tensor is even under the permutation of $q$ and $r$.
It satisfies the ``Ward identity''
\bqa
\!\!\!p_{\mu}\tilde{\cal T}^{\mu}(p,q,r)=
\tilde{\cal T}^{\mu}(q)-\tilde{\cal T}^{\mu}(r)\;,
\eqa
The electron-photon vertex therefore satisfies
the Ward identity
\bqa
p_{\mu}\Gamma^{\mu}(p,q,r)=S^{-1}(q)-S^{-1}(r)\;.
\label{qward1}
\eqa

\subsection{Electron-photon four-vertex}
We define the electron-photon four-point vertex with outgoing photon 
        momenta $p$ and $q$, incoming electron 
momentum $r$, and outgoing
        electron  momentum $s$. It reads
\bqa\nonumber
\Gamma^{\mu\nu}(p,q,r,s) &=& 
    - 2 e^2 m_f^2 \tilde{\cal T}^{\mu\nu}(p,q,r,s)  \\
&    \equiv &2 e^2 \Gamma^{\mu\nu}  \, ,
\label{4qgv}
\eqa
There is no tree-level term. The tensor in the 
HTL correction term is only defined for $p+q-r+s=0$,
\bqa\nonumber
\tilde{{\cal T}}^{\mu\nu}(p,q,r,s)
&=&\left\langle
y^{\mu}y^{\nu}\left({1\over r\!\cdot\!y}+{1\over s\!\cdot\!y}\right)
\right.\\&&\times \left.
{y\!\!\!/\over[(r-p)\!\cdot\!y]\;[(s+p)\!\cdot\!y]}
\right\rangle .
\label{T4-def}
\eqa
This tensor is symmetric in $\mu$ and $\nu$ and is traceless.
It satisfies the Ward identity:
\bqa
\!\!p_{\mu}\Gamma^{\mu\nu}(p,q,r,s)\!=\!\Gamma^{\nu}(q,r-p,s)-
\Gamma^{\nu}(q,r,s+p)\,.
\label{qward2}
\eqa

\subsection{HTL electron counterterm}
The Feynman rule for the insertion of an HTL electron counterterm into an 
electron
propagator is
\bqa
i\Sigma(p)\;,
\eqa
where $\Sigma(p)$ is the HTL electron self-energy given in~(\ref{qself}).
\subsection{Imaginary-time formalism}
\label{app:ITF}

In the imaginary-time formalism,
Minkowski energies have discrete imaginary values
$p_0 = i (2 \pi n T)$
and integrals over Minkowski space are replaced by sum-integrals over
Euclidean vectors $(2 \pi n T, {\bf p})$.
We will use the notation $P=(P_0,{\bf p})$ for Euclidean momenta.
The magnitude of the spatial momentum will be denoted $p = |{\bf p}|$,
and should not be confused with a Minkowski vector.
The inner product of two Euclidean vectors is
$P \cdot Q = P_0 Q_0 + {\bf p} \cdot {\bf q}$.
The vector that specifies the thermal rest frame
remains $n = (1,{\bf 0})$.

The Feynman rules for Minkowski space given above can be easily
adapted to Euclidean space.  The Euclidean tensor in a given
Feynman rule is obtained from the corresponding Minkowski tensor
with raised indices by replacing each Minkowski energy $p_0$
by $iP_0$, where $P_0$ is the corresponding Euclidean energy,
and multipying by $-i$ for every $0$ index.
This prescription transforms $p=(p_0,{\bf p})$ into $P=(P_0,{\bf p})$,
$g^{\mu \nu}$ into $- \delta^{\mu \nu}$,
and $p\!\cdot\!q$ into $-P\!\cdot\!Q$.
The effect on the HTL tensors defined in (\ref{T2-def}),
(\ref{T3-def}), and (\ref{T4-def}) is equivalent to
substituting $p\!\cdot\!n \to - P\!\cdot\!N$ where $N = (-i,{\bf 0})$,
$p\!\cdot\!y \to -P\!\cdot\!Y$ where $Y = (-i,{\bf \hat y})$,
and $y^\mu \to Y^\mu$.
For example, the Euclidean tensor corresponding to (\ref{T2-def}) is
\bqa
{\cal T}^{\mu\nu}(P,-P)=
\left \langle Y^{\mu}Y^{\nu}{P\!\cdot\!N \over P\!\cdot\!Y}
\right\rangle \;.
\label{T2E-def}
\eqa
%
The average is taken over the directions of the unit vector ${\bf \hat y}$.

Alternatively, one can calculate a diagram
by using the Feynman rules for Minkowski momenta,
reducing the expressions for diagrams to scalars,
and then make the appropriate substitutions,
such as $p^2 \to -P^2$, $p \cdot q \to - P \cdot Q$,
and $n \cdot p \to i n \cdot P$.
For example, the propagator functions (\ref{Delta-T:M})
and (\ref{Delta-L:M}) become
\bqa
\Delta_T(P)&=&{-1 \over P^2 + \Pi_T(P)}\;,
\label{Delta-T}
\\
\Delta_L(P)&=&{1 \over p^2+\Pi_L(P)}\;.
\label{Delta-L}
\eqa
%
The expressions for the HTL self-energy functions $\Pi_T(P)$
and $\Pi_L(P)$ are given by
(\ref{PiT-T}) and (\ref{PiT-L}) with $n_p^2$ replaced by
$n_P^2 = p^2/P^2$ and ${\cal T}^{00}(p,-p)$ replaced by
\bqa
{\cal T}_P &=& {w(\epsilon)\over2}
        \int_{-1}^1dc\;(1-c^2)^{-\epsilon}{iP_0\over iP_0-pc} \;.
\label{TP-def}
\eqa
%
Note that this function differs by a sign from the 00 component
of the Euclidean tensor corresponding
to~(\ref{T2-def}):
\bqa
{\cal T}^{00}(P,-P) = - {\cal T}^{00}(p,-p)\bigg|_{p_0 \to iP_0}
                    = - {\cal T}_P \;.
\eqa
%
A more convenient form for calculating sum-integrals
that involve the function ${\cal T}_P$ is
\bqa
{\cal T}_P &=&
        \left\langle {P_0^2 \over P_0^2 + p^2c^2} \right\ranglec \, ,
\label{TP-int}
\eqa
%
where the angular brackets represent an average over $c$ defined by
\begin{equation}
\left\langle f(c) \right\rangle_{\!c} \equiv w(\epsilon) \int_0^1 dc \,
(1-c^2)^{-\epsilon} f(c) \, ,
\label{c-average}
\end{equation}
%
and $w(\epsilon)$ is given in~(\ref{weight}).

\section{Sum-integrals}
\setcounter{equation}{0}
\label{app:sumint}

In the imaginary-time formalism for thermal field theory, 
the four-momentum $P=(P_0,{\bf p})$ is Euclidean with $P^2=P_0^2+{\bf p}^2$. 
The Euclidean energy $p_0$ has discrete values:
$P_0=2n\pi T$ for bosons and $P_0=(2n+1)\pi T$ for fermions,
where $n$ is an integer. 
Loop diagrams involve sums over $P_0$ and integrals over ${\bf p}$. 
With dimensional regularization, the integral is generalized
to $d = 3-2 \epsilon$ spatial dimensions.
We define the dimensionally regularized sum-integral by
\bqa
  \hbox{$\sum$}\!\!\!\!\!\!\int_{P} &\equiv& 
  \left(\frac{e^\gamma\mu^2}{4\pi}\right)^\epsilon\;
  T\sum_{P_0=2n\pi T}\:\int {d^{3-2\epsilon}p \over (2 \pi)^{3-2\epsilon}}\;,\\ 
  \hbox{$\sum$}\!\!\!\!\!\!\int_{\{P\}} \!\!\! &\equiv& \!\!\! 
  \left(\frac{e^\gamma\mu^2}{4\pi}\right)^\epsilon\;
  T\sum_{P_0=(2n+1)\pi T}\:\int {d^{3-2\epsilon}p \over (2 \pi)^{3-2\epsilon}}
\;,
\label{sumint-def}
\eqa
where $3-2\epsilon$ is the dimension of space and $\mu$ is an arbitrary
momentum scale. 
The factor $(e^\gamma/4\pi)^\epsilon$
is introduced so that, after minimal subtraction 
of the poles in $\epsilon$
due to ultraviolet divergences, $\mu$ coincides 
with the renormalization
scale of the $\overline{\rm MS}$ renormalization scheme.

Below we list the sum-integrals required to complete the three-loop
calculation.  We refer to~\cite{htl2,aps1} for details concerning
the sum-integral evaluations.

\subsection{One-loop sum-integrals}
The simple one-loop sum-integrals required in our calculations
can be derived from the formulas  
\bqa\nonumber
\sumint_{P}{p^{2m}\over(P^2)^n}
\!\!&=&\!\!
\left({\mu\over4\pi T}\right)^{2\epsilon}
{2\Gamma({3\over2}+m-\epsilon)\Gamma(n-{3\over2}-m+\epsilon)
\over\Gamma(n)\Gamma(2-2\epsilon)}
\\\nonumber
&&\;\times\,
\Gamma(1-\epsilon)\zeta(2n-2m-3+2\epsilon)
e^{\epsilon\gamma}
\\ &&
\;\;\;\;\times\,T^{4+2m-2n}(2\pi)^{1+2m-2n}\;,
\\
\sumint_{\{P\}}\!\!{p^{2m}\over(P^2)^n}\!\!&=&\!\!
(2^{2n-2m-d}-1)\sumint_{P}{p^{2m}\over(P^2)^n}\;. 
\eqa
\\
\noindent
The specific bosonic one-loop sum-integrals needed are

\begin{widetext}
\bqa
\sumint_{P}\log P^2&=&-{\pi^2\over45}T^4
\left[
1+{\cal O}\!\left(\epsilon\right)\right]\;,\\
\sumint_{P}{1\over P^2}
&=&{T^2\over12}
\left({\mu\over4\pi T}\right)^{2\epsilon}
\Bigg[1+\left(2+2{\zeta^{\prime}(-1)\over\zeta(-1)}
\right)\epsilon
+{\cal O}\!\left(\epsilon^2\right)\Bigg]\;, 
\\ 
\sumint_P {1 \over (P^2)^2} &=&
{1 \over (4\pi)^2} \left({\mu\over4\pi T}\right)^{2\epsilon} 
\left[ {1 \over \epsilon} + 2 \gamma
+{\cal O}\!\left(\epsilon\right)
\right] \;,
\\ 
\sumint_P {1 \over p^2 P^2} &=&
{1 \over (4\pi)^2} \left({\mu\over4\pi T}\right)^{2\epsilon} 
2 \left[ {1\over\epsilon} + 2 \gamma + 2 
+{\cal O}\!\left(\epsilon\right)
\right]\;.
 \label{ex1}
\eqa

The specific fermionic one-loop sum-integrals needed are
\bqa
\sumint_{\{P\}}\log P^2&=&{7\pi^2\over360}T^4
\left[1+{\cal O}\!\left(\epsilon\right)
\right]
\;,\\ 
\sumint_{\{P\}}{1\over P^2}
&=&-{T^2\over24}
\left({\mu\over4\pi T}\right)^{2\epsilon}
\bigg[1
+\bigg(
2-2\log2
+2{\zeta^{\prime}(-1)\over\zeta(-1)}
\bigg)\epsilon 
+{\cal O}\!\left(\epsilon^2\right)
\bigg]\;,
\label{simple1}
\\ 
\sumint_{\{P\}}{1\over(P^2)^2}&=&
{1\over(4\pi)^2}\left({\mu\over4\pi T}\right)^{2\epsilon}
\left[ {1 \over \epsilon} + 2 \gamma+4\log2
+{\cal O}\!\left(\epsilon\right)
\right] ,
\\ 
\sumint_{\{P\}}{p^2\over(P^2)^2}&=&-{T^2\over16}
\left({\mu\over4\pi T}\right)^{2\epsilon}
\bigg[1
+\bigg({4\over3}-2\log2
+2{\zeta^{\prime}(-1)\over\zeta(-1)}\bigg)\epsilon
+{\cal O}\!\left(\epsilon^2\right)
\bigg]\;,
\\ 
\sumint_{\{P\}}{p^2\over(P^2)^3}&=&{1\over(4\pi)^2}
\left({\mu\over4\pi T}\right)^{2\epsilon}
\, {3\over4}
\bigg[{1\over\epsilon}+2\gamma-{2\over3}
+4\log2
+{\cal O}\!\left(\epsilon\right)
\bigg] ,\\ 
\sumint_{\{P\}}{p^4\over(P^2)^4}&=&{1\over(4\pi)^2}
\left({\mu\over4\pi T}\right)^{2\epsilon}
{5\over8}\bigg[
{1\over\epsilon}+2\gamma-{16\over15}
+4\log2
+{\cal O}\!\left(\epsilon\right)
\bigg]
 \;,
\\ 
\sumint_{\{P\}}{1\over p^2P^2}&=&{1\over(4\pi)^2}
\left({\mu\over4\pi T}\right)^{2\epsilon}
2\bigg[
{1\over\epsilon}+2+2\gamma+4\log2
+{\cal O}\!\left(\epsilon\right)
\bigg]\;.
\eqa

The errors are all of one order higher in $\epsilon$  than 
the smallest term shown.
The number $\gamma_1$ is the first Stieltjes gamma constant
defined by the equation
\begin{equation}
\label{zeta}
\zeta(1+z) = {1 \over z} + \gamma - \gamma_1 z + O(z^2)\;.
\end{equation}

We also need some more difficult one-loop sum-integrals 
that involve the HTL function 
defined in (\ref{def-tf}). 
The specific bosonic sum-integrals needed are
\bqa
\sumint_P {1 \over p^4} {\cal T}_P &=&
{1 \over (4\pi)^2} \left({\mu\over4\pi T}\right)^{2\epsilon}
(-1)\left[ 
{1 \over \epsilon} + 2 \gamma 
+ 2\log2
+{\cal O}\!\left(\epsilon\right)
\right]
\;,
\label{exa}
\\ 
\sumint_P {1 \over p^2 P^2} {\cal T}_P &=&
{1 \over (4\pi)^2} \left({\mu\over4\pi T}\right)^{2\epsilon}
\left[ 
2 \log2 \left({1 \over \epsilon} + 2 \gamma \right)
+ 2 \log^2 2 + {\pi^2 \over 3}  
+{\cal O}\!\left(\epsilon\right)
\right] \;, \\
\sumint_P {1 \over p^4} ({\cal T}_P)^2 &=&
{1 \over (4\pi)^2} \left({\mu\over4\pi T}\right)^{2\epsilon}
\left( - {2 \over 3} \right)
\left[ (1+ 2 \log 2) \left( {1 \over \epsilon} + 2 \gamma \right)
       - {4 \over 3} + {22 \over 3} \log 2 + 2 \log^2 2 
+{\cal O}\!\left(\epsilon\right)
\right] 
\;.
\label{sumint-T:5}
\eqa
%
The specific fermionic sum-integrals needed are
\bqa
\sumint_{\{P\}} {1 \over (P^2)^2} {\cal T}_P &=&
{1 \over (4\pi)^2} \left({\mu\over4\pi T}\right)^{2\epsilon}
{1 \over 2}
\bigg[ {1 \over \epsilon} + 2 \gamma + 1 
+ 4\log2
+{\cal O}\!\left(\epsilon\right)
\bigg] \;,
\label{ht1} \\ 
\sumint_{\{P\}}{1\over p^2P^2}{\cal T}_P&=&
{1\over(4\pi)^2}\left({\mu\over4\pi T}\right)^{2\epsilon}
\left[ 2\log2 \left({1 \over \epsilon} + 2 \gamma \right) 
        + 10 \log^22 + {\pi^2 \over 3}  
+{\cal O}\!\left(\epsilon\right)\right]\;,
\\ 
\sumint_{\{P\}}{1\over P^2P_0^2}{\cal T}_P&=&
{1\over(4\pi)^2}\left({\mu\over4\pi T}\right)^{2\epsilon}
\Bigg[ {1\over \epsilon^2}
+2(\gamma+2\log2){1\over \epsilon}
+{\pi^2\over4}
+4\log^22
+8\gamma\log2-4\gamma_1
+{\cal O}\!\left(\epsilon\right)\bigg]\;,
\\ 
\sumint_{\{P\}}{1\over p^2P_0^2}\left({\cal T}_P\right)^2&=&
{4\over(4\pi)^2}\left({\mu\over4\pi T}\right)^{2\epsilon}
\bigg[\log2\left({1\over\epsilon}+2\gamma\right)
+5\log^22
+{\cal O}\!\left(\epsilon\right)\bigg]\;,
\label{htlf} \\
\sumint_{\{P\}}{1\over P^2}
\bigg\langle {1\over(P\!\cdot\!Y)^2} \bigg\rangle_{\hat{\bf y}}
&=&
{1\over(4\pi)^2}\left({\mu\over4\pi T}\right)^{2\epsilon} 
(-1)\bigg[{1\over\epsilon}-1+2\gamma
+4\log2 
+{\cal O}\!\left(\epsilon\right)
\bigg]\;.
\label{cp1ly} 
\eqa

\subsection{Two-loop sum-integrals}

The simple two-loop sum-integrals that are needed are
\bqa
\sumint_{\{PQ\}}{1\over P^2Q^2R^2}&=&0 \; , \\ 
\sumint_{\{PQ\}}{1\over P^2Q^2r^2}&=&
{T^2 \over (4 \pi)^2} \left({\mu\over4\pi T}\right)^{4\epsilon}
\left(-{1\over6}\right)\left[{1\over\epsilon}
+4
-2\log2+4{\zeta^{\prime}(-1)\over\zeta(-1)}
+{\cal O}\!\left(\epsilon\right)
\right] 
\label{two1}
\; , \\ 
\sumint_{\{PQ\}}{q^2\over P^2Q^2r^4}&=&
{T^2 \over (4 \pi)^2} \left({\mu\over4\pi T}\right)^{4\epsilon} 
\left(-{1\over12}\right)\left[{1\over\epsilon}+
{11\over3}
+2\gamma-2\log2+2{\zeta^{\prime}(-1)\over\zeta(-1)}
+{\cal O}\!\left(\epsilon\right)
\right] , 
\label{two2}
\\ 
\sumint_{\{PQ\}}{q^2\over P^2Q^2r^2R^2}&=&
{T^2 \over (4 \pi)^2} \left({\mu\over4\pi T}\right)^{4\epsilon}
\left(-{1\over72}\right)
\bigg[{1\over\epsilon}
-7.00164
+{\cal O}\!\left(\epsilon\right)
\bigg]  
\label{two3}
\;,\\ 
\sumint_{\{PQ\}}{P\!\cdot\!Q\over P^2Q^2r^4}&=&
{T^2 \over (4 \pi)^2} \left({\mu\over4\pi T}\right)^{4\epsilon}
\left(-{1\over36}\right)
\left[1-6\gamma+6{\zeta^{\prime}(-1)\over\zeta(-1)}
+{\cal O}\!\left(\epsilon\right)
\right] 
\label{twolast}
\;, \\ 
\sumint_{\{PQ\}}{p^2\over q^2P^2Q^2R^2}&=&
{T^2\over(4\pi)^2}\left({\mu\over4\pi T}\right)^{4\epsilon}
\left({5\over72}\right)
\left[
{1\over\epsilon}
+9.55216
+{\cal O}\!\left(\epsilon\right)
\right] 
\label{ntwo1}
\;,\\ 
\sumint_{\{PQ\}}{r^2\over q^2P^2Q^2R^2}&=&
{T^2\over(4\pi)^2}\left({\mu\over4\pi T}\right)^{4\epsilon}
\left(-{1\over18}\right)
\left[
{1\over\epsilon}
+8.14234
+{\cal O}\!\left(\epsilon\right)
\right] , 
\label{ntwo2}
\eqa
%
where $R=-(P+Q)$ and $r=|{\bf p}+{\bf q}|$.  The corrections are all of order
$\epsilon^2$. 
We also need some more difficult two-loop sum-integrals
that involve the functions ${\cal T}_P$ defined in (\ref{def-tf}),

\bqa
\sumint_{\{PQ\}}{1\over P^2Q^2r^2}{\cal T}_R&=&
{T^2 \over (4 \pi)^2} \left({\mu\over4\pi T}\right)^{4\epsilon}
	\left(-{1\over48}\right)
	\Bigg[{1\over\epsilon^2} + \left( 2 +12\log 2 + 4 {\zeta'(-1) 
\over \zeta(-1)} \right)
	{1\over\epsilon} +136.362 \,\, 
+{\cal O}\!\left(\epsilon\right)
\Bigg]\,, 
\label{htlf1}
\\ 
\sumint_{\{PQ\}} {q^2\over P^2Q^2r^4}{\cal T}_R &=& 
{T^2 \over (4 \pi)^2} \left({\mu\over4\pi T}\right)^{4\epsilon} 
\left(-{1\over576}\right)
	\Bigg[{1\over\epsilon^2} 
	+\left({26\over3}+52\log2+4{\zeta'(-1) \over \zeta(-1)}\right)
{1\over\epsilon}
	+ 446.412 
+{\cal O}\!\left(\epsilon\right)
\!\Bigg]\!, 
\label{htlf2}
\\ 
\sumint_{\{PQ\}}{P\!\cdot\!Q\over P^2Q^2r^4}{\cal T}_R &=&
	{T^2 \over (4 \pi)^2} \left({\mu\over4\pi T}\right)^{4\epsilon} 
\left(-{1\over96}\right)
	\Bigg[{1\over\epsilon^2} +\left(4\log2+4{\zeta'(-1)\over\zeta(-1)} 
\right) 
	{1\over\epsilon} +69.174 
+{\cal O}\!\left(\epsilon\right)
\Bigg] \; , 
\label{htl3}
\\ 
\sumint_{\{PQ\}}{r^2-p^2\over P^2q^2Q^2_0R^2}{\cal T}_Q&=&
	-{T^2\over(4\pi)^2}\left({\mu\over4\pi T}\right)^{4\epsilon}{1\over8}
	\Bigg[{1\over\epsilon^2} +\left(2+2\gamma +{10\over3}\log2
	+2{\zeta^{\prime}(-1)\over\zeta(-1)}\right){1\over\epsilon}
	+46.8757 
+{\cal O}\!\left(\epsilon\right)
\Bigg] \; . 
\label{com2l}
\eqa

\subsection{Three-loop sum-integrals}

\noindent
The three-loop sum-integrals needed are
\bqa
\sumint_{PQR}{1\over P^2Q^2R^2(P+Q+R)^2}&=&
{1\over(4\pi)^2}\left({T^2\over12}\right)^2
\left({\mu\over4\pi T}\right)^{6\epsilon}
\left[{6\over\epsilon}+{182\over5}
-12{\zeta^{\prime}(-3)\over\zeta(-3)}
+48{\zeta^{\prime}(-1)\over\zeta(-1)}
+{\cal O}\!\left(\epsilon\right)
\right],\\ \nonumber 
\sumint_{\{PQR\}}{1\over P^2Q^2R^2(P+Q+R)^2}&=&
{1\over(4\pi)^2}\left({T^2\over12}\right)^2
\left({\mu\over4\pi T}\right)^{6\epsilon}
\left[{3\over2\epsilon}+{173\over20}
-{63\over5}\log2
-3{\zeta^{\prime}(-3)\over\zeta(-3)}
\right.\\ &&\left.
+12{\zeta^{\prime}(-1)\over\zeta(-1)}
+{\cal O}\!\left(\epsilon\right)
\right]\;,\\
\nonumber
\sumint_{PQ\{R\}}{1\over P^2Q^2R^2(P+Q+R)^2}&=&
{1\over(4\pi)^2}\left({T^2\over12}\right)^2
\left({\mu\over4\pi T}\right)^{6\epsilon}
\left[-{3\over4\epsilon}-{179\over40}
+{51\over10}\log2
+{3\over2}{\zeta^{\prime}(-3)\over\zeta(-3)}
-6{\zeta^{\prime}(-1)\over\zeta(-1)} \right. \nonumber \\
&&
+ {\cal O}\!\left(\epsilon\right)
\biggr]\;,\\ 
\nonumber
\sumint_{\{P\}QR}{Q\!\cdot\!R\over P^2Q^2R^2(P+Q)^2(P+R)^2}&=&
{1\over(4\pi)^2}\left({T^2\over12}\right)^2
\left({\mu\over4\pi T}\right)^{6\epsilon}
\left[{3\over8\epsilon}+{9\over4}\gamma+{361\over160}
+{57\over10}\log2
+{3\over2}{\zeta^{\prime}(-3)\over\zeta(-3)}
\right.\\&&\left.
-{3\over2}{\zeta^{\prime}(-1)\over\zeta(-1)}
+{\cal O}\!\left(\epsilon\right)
\right]\;, \\
\nonumber
\sumint_{P\{QR\}}{(Q\!\cdot\!R)^2\over P^2Q^2R^2(P+Q)^2(P+R)^2}&=&
{1\over(4\pi)^2}\left({T^2\over12}\right)^2
\left({\mu\over4\pi T}\right)^{6\epsilon}
\left[{5\over24\epsilon}+{1\over4}\gamma
+{23\over24}
-{8\over5}\log2
-{1\over6}{\zeta^{\prime}(-3)\over\zeta(-3)}
\right.\\&&\left.
+{7\over6}{\zeta^{\prime}(-1)\over\zeta(-1)}
+{\cal O}\!\left(\epsilon\right)
\right]\;.
\eqa
The three-loop sum-integrals were first calculated by Arnold and Zhai,
and calculational details can be found in Ref.~\cite{AZ-95}.

\section{Three-dimensional integrals}\label{appb}
Dimensional regularization can be used to
regularize both the ultraviolet divergences and infrared divergences
in three-dimensional integrals over momenta.
The spatial dimension is generalized to  $d = 3-2\epsilon$ dimensions.
Integrals are evaluated at a value of $d$, for which they converge, and then
analytically continued to $d=3$.
We use the integration measure
\begin{equation}
 \int_p\;\equiv\;
  \left(\frac{e^\gamma\mu^2}{4\pi}\right)^\epsilon\;
\:\int {d^{3-2\epsilon}p \over (2 \pi)^{3-2\epsilon}}\;.
\label{int-def}
\end{equation}


\noindent
The one-loop integrals needed are of the form
\bqa\nonumber
I_n&\equiv&\int_p{1\over(p^2+m^2)^n}\\
&=&{1\over8\pi}(e^{\gamma_E}\mu^2)^{\epsilon}
{\Gamma(n-\mbox{$3\over2$}+\epsilon)
\over\Gamma(\mbox{$1\over2$})
\Gamma(n)}m^{3-2n-2\epsilon}
\;.
\eqa
Specifically, we need
\bqa\nonumber
I_0^{\prime}&\equiv&
\int_p\log(p^2+m^2)\\
&=&
-{m^3\over6\pi}\left({\mu\over2m}\right)^{2\epsilon}
\left[
1
+{8\over3}
\epsilon
+{\cal O}\!\left(\epsilon^2\right)
\right]\;,\\ 
I_1&=&-{m\over4\pi}\left({\mu\over2m}\right)^{2\epsilon}
\left[
1+2
\epsilon
+{\cal O}\!\left(\epsilon^2\right)
\right]\;,\\
\label{i2}
I_2&=&{1\over8\pi m}\left({\mu\over2m}\right)^{2\epsilon}
\left[
1
+{\cal O}\!\left(\epsilon\right)
\right]
\;.
\eqa
\end{widetext}

\end{document}